\documentclass[draft]{agujournal2019}
\usepackage{url} %this package should fix any errors with URLs in refs.
\usepackage{lineno}
\usepackage[inline]{trackchanges} %for better track changes. finalnew option will compile document with changes incorporated.
\usepackage{soul}
\usepackage{mathtools}   % loads »amsmath«
\draftfalse

\journalname{JGR: Oceans}

\let\oldequation\equation
\let\oldendequation\endequation

\renewenvironment{equation}
{\linenomathNonumbers\oldequation}
{\oldendequation\endlinenomath}

\begin{document}

%% ------------------------------------------------------------------------ %%
%  Title
%
% (A title should be specific, informative, and brief. Use
% abbreviations only if they are defined in the abstract. Titles that
% start with general keywords then specific terms are optimized in
% searches)
%
%% ------------------------------------------------------------------------ %%

% Example: \title{This is a test title}

\title{Quantification of salt transports due to exchange flow and tidal flow in estuaries}

%% ------------------------------------------------------------------------ %%
%
%  AUTHORS AND AFFILIATIONS
%
%% ------------------------------------------------------------------------ %%

% Authors are individuals who have significantly contributed to the
% research and preparation of the article. Group authors are allowed, if
% each author in the group is separately identified in an appendix.)

% List authors by first name or initial followed by last name and
% separated by commas. Use \affil{} to number affiliations, and
% \thanks{} for author notes.
% Additional author notes should be indicated with \thanks{} (for
% example, for current addresses).

% Example: \authors{A. B. Author\affil{1}\thanks{Current address, Antartica}, B. C. Author\affil{2,3}, and D. E.
% Author\affil{3,4}\thanks{Also funded by Monsanto.}}

\authors{Bouke Biemond\affil{1} , Huib E. de Swart\affil{1}, Henk A. Dijkstra \affil{1} }%Manuel D{\'\i}ez-Minguito \affil{2} }

% \affiliation{1}{First Affiliation}
% \affiliation{2}{Second Affiliation}
% \affiliation{3}{Third Affiliation}
% \affiliation{4}{Fourth Affiliation}

\affiliation{1}{Institute for Marine and Atmospheric research Utrecht, Department of Physics, 
	Utrecht University, Utrecht, the Netherlands.}
%\affiliation{2}{Environmental Fluid Dynamics Group, Andalusian Institute for Earth System Research, University of Granada, Granada, Spain}

%(repeat as many times as is necessary)

%% Corresponding Author:
% Corresponding author mailing address and e-mail address:

% (include name and email addresses of the corresponding author.  More
% than one corresponding author is allowed in this LaTeX file and for
% publication; but only one corresponding author is allowed in our
% editorial system.)

% Example: \correspondingauthor{First and Last Name}{email@address.edu}

\correspondingauthor{Bouke Biemond}{w.t.biemond@uu.nl}

%% Keypoints, final entry on title page.

%  List up to three key points (at least one is required)
%  Key Points summarize the main points and conclusions of the article
%  Each must be 140 characters or fewer with no special characters or punctuation and must be complete sentences

% Example:
% \begin{keypoints}
% \item	List up to three key points (at least one is required)
% \item	Key Points summarize the main points and conclusions of the article
% \item	Each must be 140 characters or fewer with no special characters or punctuation and must be complete sentences
% \end{keypoints}

\begin{keypoints}
\item Salt intrusion in estuaries due to exchange flow and tidal flow is quantified with an idealized semi-analytical model.
\item Explicit expressions are derived that measure the importance of different salt intrusion mechanisms. 
\item Net salt transport by the depth-averaged tidal flow is larger than the transport associated with the vertical shear of the tidal flow. 
\end{keypoints}

%% ------------------------------------------------------------------------ %%
%
%  ABSTRACT and PLAIN LANGUAGE SUMMARY
%
% A good Abstract will begin with a short description of the problem
% being addressed, briefly describe the new data or analyses, then
% briefly states the main conclusion(s) and how they are supported and
% uncertainties.

% The Plain Language Summary should be written for a broad audience,
% including journalists and the science-interested public, that will not have 
% a background in your field.
%
% A Plain Language Summary is required in GRL, JGR: Planets, JGR: Biogeosciences,
% JGR: Oceans, G-Cubed, Reviews of Geophysics, and JAMES.
% see http://sharingscience.agu.org/creating-plain-language-summary/)
%
%% ------------------------------------------------------------------------ %%

%% \begin{abstract} starts the second page

\begin{abstract}
	To understand mechanisms of salt intrusion in estuaries, we develop a semi-analytical model, that explicitly accounts for salt transport by both exchange flow and tidal flow. This model, after calibration, successfully hindcasts hydrodynamics and salinity dynamics in three estuaries that have strongly different characteristics. 
	We find, from analyzing the model results for these three estuaries, that salt transport processes by exchange flow and tidal flow interact through the subtidal stratification. Transport by exchange flow creates stratification, thereby generating a phase shift of tidal salinity with respect to the tidal flow, which is important for the magnitude of the tidal salt transport. Conversely, the strength of tidal currents determines the vertical mixing that breaks down stratification. 
	A new analytical formulation is presented for the component of the salt transport driven by the depth-averaged tidal flow. 
	This salt transport is larger than the component associated with the vertical shear of the tidal current. 
	Finally, a method that yields analytical equations that quantify the importance of different contributions to the salt transport using only primary information is developed using approximate solutions for the subtidal stratification. This method performs well for the estuaries considered.	
\end{abstract}

\section*{Plain Language Summary}

The salinity distribution in an estuary, the transition area between river and sea, is determined by different salt transport mechanisms. Landward currents into the estuary will increase the extent of the salt intrusion, and seaward flows will decrease the extent. 
Currents in estuaries are driven by different forcing agents. Discharge from the river causes a seaward current. The fact that salty water is denser than fresh water drives landward currents near the bottom, but seaward currents near the surface, a so-called exchange flow. Tidal currents, driven by water level variations at the estuary mouth, alternate their direction every tidal cycle. 
In this study, we develop a model to quantify the salt transport associated with these different flow types in different estuaries. After having confirmed that the model results agree well with measurements, we show that the exchange flow is the most important driver of salt import in the Delaware Estuary and the Loire Estuary, but salt import by the tidal currents is more important in the Guadalquivir Estuary. We provide explanations for the strength of these different salt transport components and their interaction. 

 % ----------------------------%
\section{Introduction}
\label{sec:intr5}
% ----------------------------%

The longitudinal distribution of subtidal salinity in estuaries is determined by the competition between downstream and upstream net, i.e. tidally averaged, salt transport. The major driver of downstream transport is river discharge. Upstream transport results, among others, from exchange flow and tidal flow (\citeA{valle2010}, and references therein). 
The importance of these different drivers is determined by environmental conditions, e.g. the amount of freshwater input, tidal strength, magnitude and direction of the wind, and the estuarine geometry. A complication is that these conditions differ per estuary, per location in an estuary and in time. 
The aim of the present study is to increase understanding of the physical processes that determine upstream salt transport, i.e. those responsible for salt intrusion in estuaries. 

It was demonstrated by \citeA{dyer1973estuaries} and \citeA{fischer1976} that net longitudinal salt transport due to different mechanisms can be computed if vertical profiles of currents and salinity are available during one or more tidal cycles. 
From analysis of a 25-hour time series at a section of the Hudson River Estuary, \citeA{hunkins81} concluded that upstream transport due to the exchange flow is dominant in this estuary. But when using a 70-day moored time series, the ratio between salt transport associated with the exchange flow and tidal flow was found to be 70\%/30\% \cite{bowen2003}. A third study in this estuary, using a cross-channel array of moored current profilers and salinity sensors combined with ship surveys, reported a ratio of about 90\%/10\% \cite{lerczak06}.
Using field data collected in the Guadalquivir Estuary, \citeA{diez2013spa} found that tidal flow contributes most to upstream salt transport. 
In contrast, in the Delaware Estuary, exchange flow is overall the strongest actor in upstream salt transport, but the relative contribution from transport due to tidal flow can locally be substantial \cite{aristizabal15}. 
These observations show that various ratios of the contributions to salt transport are present in nature, even in the same estuary. To understand why this is the case, knowledge about the fundamental properties of salt intrusion processes is required. 

To quantify salt transport and salt intrusion, idealized semi-analytical models are helpful tools. This is because such models are flexible in design, they are computationally fast (so suitable for sensitivity studies) and explicit equations can be derived for different components of the flow and salinity. This makes it easy to separate the salt transport into contributions from different agents, and thereby gain insight into the dynamics. 
The early idealized model of \citeA{hansen1965v} represented longitudinal salt transport by river flow and exchange flow as an advective process and transport by tidal flow as a dispersive process. Using this framework, \citeA{hansen1966} present a diagram, in which estuaries are classified in terms of two dimensionless parameters. The first measures the strength of the freshwater discharge, which determines the seaward transport of salt. The second parameter denotes the strength of the vertical mixing, which controls the strength of the upstream transport due to the exchange flow.
To compute the longitudinal salinity profile in this model analytically, \citeA{chatwin1976} assumed that the tidal salt transport is negligible  (see also \citeA{Chatwin1975}), and justified this a posteriori by comparing the computed salinity with observations. 
Later studies \cite{MC04,MC07} present semi-analytical solutions for the longitudinal salinity profile in the \citeA{hansen1965v} model when also accounting for horizontal dispersive salt transport caused by tides. With this, updated versions of the classification diagram were presented \cite{guha13,GMC14}. However, finding a suitable dispersion coefficient to model tidal salt transport as horizontal dispersion is far from straightforward, and \citeA{MC07} mentions that `this parameter is very poorly constrained'.

Idealized models which explicitly solve for advective salt transport (defined as the cross-sectionally integrated product of normal velocity and salinity) by tides were presented by \citeA{mccarthy1993residual} and \citeA{wei2016}. They showed that the tidal advective transport acts like a dispersive process, with a dispersion coefficient that depends on the local characteristics of the tidal flow. Those studies, however, assume well-mixed conditions in the vertical, with the consequence that advection of salt by the exchange flow is excluded. 

None of the models mentioned so far are suitable to simultaneously describe advective salt transport by both exchange flow and tidal flow and their interaction. However, the field observations mentioned above, but also numerical models (e.g. \citeA{gong2014,maccready24}), clearly indicate that such a model is of added value to be able to gain deeper understanding of the various drivers of upstream salt transport in an estuary.
Moreover, exchange flow and tidal flow are intrinsically connected to each other. 
Turbulence generated by tidal currents is important to the vertical structure of velocity and salinity in estuaries \cite{Geyer2000,Stacey2001}, which determines the strength of the salt transport by exchange flow. 
On the other hand, the `moving-plane' coordinate system introduced by \citeA{Dronkers1986} shows that the strength of the tidal transport is determined by the interaction of the tidally varying salinity with shear-related mechanisms, such as the exchange flow \cite{Garcia2023}. 
Fundamental knowledge of salt advection by both exchange flow and tidal flow is therefore essential to understand salt transport in estuaries.

Here, we present a model to investigate these phenomena, and apply it to three estuaries: the Delaware Estuary, the Guadalquivir Estuary and the Loire Estuary. 
The specific aims of this study are: 
1) Demonstrate that our model is capable of representing the gross hydrodynamic and salinity characteristics of different estuaries. 
2) Quantify the contributions of different salt transport components, including their interactions, to the upstream salt transport for different estuaries.
3) Provide an explanation for the magnitude and interaction of these salt transport components in terms of physical mechanisms. 
4) Estimate the relative importance of the salt transport components using only primary information, that is, information about the geometry, discharge and tidal water level variation at the mouth.

This manuscript is organized as follows. Section 2 presents the definition of salt transport and the new model. Moreover, it describes the field data used in this study. Section 3 presents results for the first two research aims. Section 4 presents results regarding the last two research aims, followed by a discussion in
Section 5 and the conclusions in Section 6.

% ----------------------------%
\section{Material}
\label{sec:meth}
% ----------------------------%

\subsection{Net salt transport}

The longitudinal subtidal salt transport $T$ in an estuarine channel, if we assume the lateral boundaries to be vertically straight, is defined as (see e.g. \citeA{fischer1976,lerczak06})
\begin{equation} \label{eq:Tdef}
T = b \left( (H+\eta) \overline{\left(u s - K_h \frac{\partial s}{\partial x} \right)}\right)_{st} .
\end{equation}
In this expression, $b$ and $H$ are the width and mean depth of the channel, respectively, $\eta$ is water level, $u$ is along-channel velocity, $s$ is salinity and $K_h$ is a horizontal dispersion coefficient. The horizontal coordinate is denoted by $x$. The subscript $st$ (subtidal) indicates that an average is taken over the tidal cycle and the bar indicates an average over depth. 
Here, it is assumed that the variables do not vary over the cross-channel direction. 
The term containing $\overline{u s}$ is the advective contribution to the salt transport, and $- K_h \frac{\partial \bar s}{\partial x}$, the term containg the dispersive contribution, represents the salt transport processes originating from horizontal mixing. 
To identify different salt transport components, we follow \citeA{Lewis1983} (see also e.g. \citeA{lerczak06} and \citeA{diez2013spa}) and define 
\begin{equation} \label{eq:dec}
	u = \bar u_{st} + u_{st}^\prime + \bar u_{ti}+ u_{ti}^\prime , \quad s = \bar s_{st} + s_{st}^\prime + \bar s_{ti} + s_{ti}^\prime ,	\quad \eta = \eta_{st} + \eta_{ti}, 
\end{equation}
in which subscript $ti$ (tidal) indicates tidally varying parameters and a prime indicates the deviation from the depth-averaged value. 
We insert this decomposition in Eq.~\ref{eq:Tdef}, and perform three approximations: variations of the water level are assumed to be small with respect to the mean water depth, variations of salinity in the tidal cycle are small with respect to the subtidal salinity, and subtidal currents are weak with respect to tidal currents.
This yields for the salt transport (see \ref{app:T} for the derivation)
\begin{equation} \label{eq:Tdec}
T =  \underbrace{\vphantom{\frac{1}{1}}b H \bar u_{st} \bar s_{st} }_{T_{Q}}+\underbrace{\vphantom{\frac{1}{1}} b H \overline{u'_{st}s'_{st}} }_{T_{e}} + \underbrace{\vphantom{\frac{1}{1}}b H (\bar u_{ti} \bar s_{ti}) _{st}}_{T_{\overline{ti}}} + \underbrace{\vphantom{\frac{1}{1}}b H \overline{(u_{ti}^\prime s_{ti}^\prime)} _{st}}_{T_{ti^\prime}} - \underbrace{\vphantom{\frac{1}{1}} b H K_h \frac{\partial \bar s_{st}}{\partial x} }_{T_{d}} . 
\end{equation}
Here, $T_{Q}$ is the salt transport due to river discharge, $T_{e}$ is the salt transport by the exchange flow, which in our model context consist of vertical shear of the river flow and density-driven flow, $T_{\overline{ti}}$ is the salt transport due to time correlation of depth-averaged tidal flow and depth-averaged salinity, $T_{ti^\prime}$ is the salt transport due to correlations in time between the departures of the tidal flow and salinity from their depth-averaged values, and $T_d$ is the salt transport due to horizontal mixing, which mainly represents unresolved processes like lateral shear dispersion \cite{Scully2007, Scully2012}. 
Note that a number of earlier studies \cite{lerczak06,aristizabal15} consider both components of the salt transport due to tidal flow as a single component. However, we will show that these components have different characteristics and therefore consider them separately. 
In the following, we will present a model to calculate these different contributions explicitly.

\subsection{Model }
\subsubsection{Domain}

\begin{figure}[h!]
	\centering
	\includegraphics[width=\textwidth]{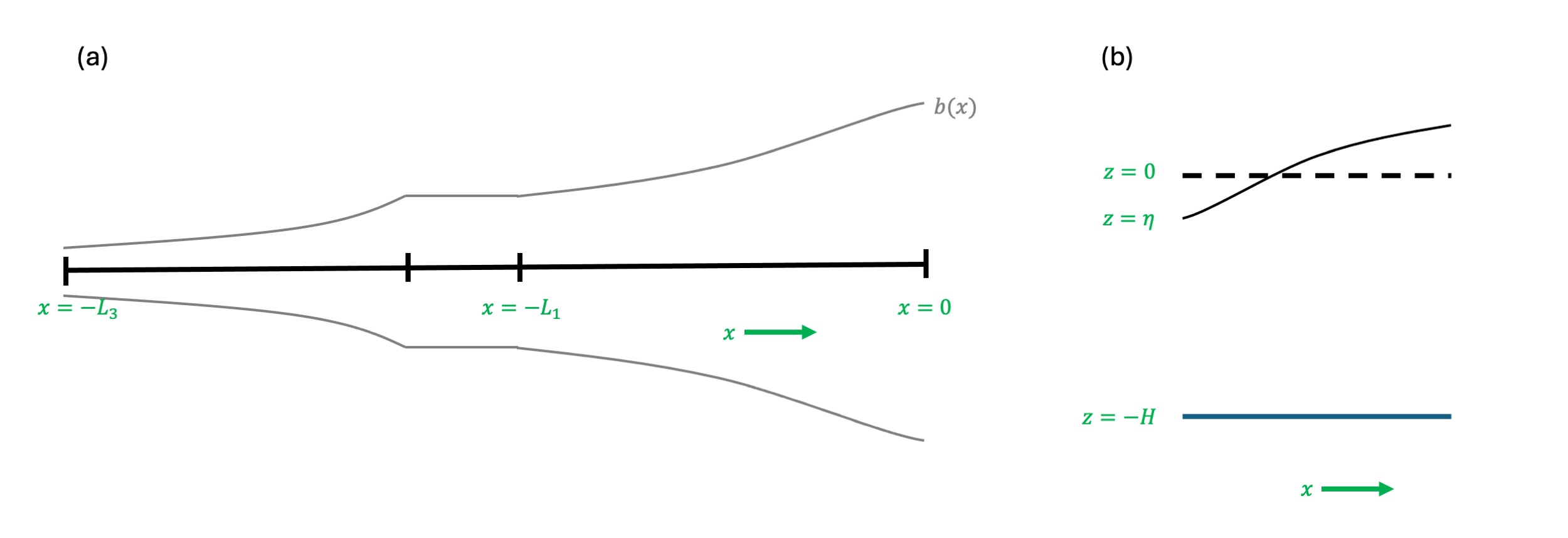}
	\caption{(a) Sketch (topview) of a typical estuarine domain. The longitudinal coordinate is denoted by $x$, width is indicated by $b$ and the black line is the $x$-axis, and the vertical bars indicate the boundaries of the segments. (b) Sideview of the model domain, where $z$ is the vertical coordinate, $H$ is the undisturbed water depth and $\eta$ is the water level. }
	\label{fig:dom}
\end{figure}

The model domain, of which a typical example is shown in Fig. \ref{fig:dom}, is an estuarine channel that consists of multiple segments. 
The depths $H_m$ of the segments in the estuary are assumed to be constant, but different segments can have a different depth.
A Cartesian coordinate frame is chosen, where the $x$-axis points from river to sea and the $z$-axis points vertically upward. The origin is the estuary mouth $x=0$ and undisturbed water surface $z=0$. The location of the river mouth is at $x=-L_{M}$, where $M$ is the number of segments. 
The width in each segment $[-L_{m+1}, -L_m]$ is described as 
\begin{equation}
	b(x) = b(-L_{m}) e^{(x+L_{m})/L_{b,m}} , % \quad \text{where} \quad -L_{m+1}<x<-L_{m},
\end{equation}
 and $L_{b,m}$ is the $e$-folding length scale which controls the width convergence of segment $m$. 
The width is continuous at the boundaries of the segments. For clarity, the subscript $m$ will be left out from now, and the presented equations will be valid for the inner part of the segments, while at their boundaries continuity conditions are imposed. 

\subsubsection{Hydrodynamics}

To find the various components of the net salt transport in Eq.~\ref{eq:Tdec}, expressions have to be found for the four components of the horizontal velocity that are defined in Eq.~\ref{eq:dec}. 
Regarding subtidal flow, we apply a similar procedure as \citeA{hansen1965v,dyer1973estuaries} and \citeA{MC04}, and construct solutions for $\bar u_{st}$ and $u'_{st}$ by solving the dominant subtidal momentum and mass balances
\begin{subequations} \label{eq:hyst}
	\begin{eqnarray} 
		&&0 = - g\frac{\partial \eta_{st}}{\partial x} + g \beta z \frac{\partial \bar s_{st}}{\partial x}  + A_{v,st} \frac{\partial^2 u_{st}}{\partial z^2} , \\
		&&\frac{1}{b} \frac{\partial }{\partial x} \left(b u_{st}\right) + \frac{\partial w_{st}}{\partial z} = 0 .
	\end{eqnarray}
\end{subequations} 
Here, $w_{st}$ is subtidal vertical velocity, $\beta= 7.6 \ 10^{-4}$ (g/kg)$^{-1}$ is the isohaline contraction coefficient, $g=9.81$ m s$^{-2}$, and $A_{v,st}$ is vertical viscosity for subtidal flow, assumed constant in time and space. These expressions describe river flow and exchange flow; tidally rectified flow is not described as the net salt transport associated with this flow is shown to be negligible (see \ref{app:T}). Note that to construct a consistent model framework, we also decomposed viscosity and vertical velocity in a subtidal and a tidally varying part, and will do the same for friction coefficients and diffusivity. The boundary conditions are: at the river boundary, a discharge is prescribed, at the boundaries of the segments discharge and water level is continuous, at the surface free slip is used and at the bottom a partial slip condition is employed, which reads
\begin{equation} \label{eq:slip}
	A_{v,st}\frac{\partial u_{st}}{\partial z} = S_{f,st} u_{st} \quad \text{with} \quad S_{f,st} = \frac{2 A_{v,st}}{H},
\end{equation}
in which $S_{f,st}$ is the friction coefficient acting on the subtidal flow. For $w_{st}$, kinematic boundary conditions are employed at the bottom and surface. From this, analytical expressions for water level $\eta_{st}$ and velocity $\bar u_{st}$ and $u'_{st}$ as a function of $\frac{\partial \bar s_{st}}{\partial x}$ are found. 
The solution procedure is detailed in \ref{app:vel}. A closure for $A_{v,st}$ will follow. 

Expressions for $\bar u_{ti}$ and $u'_{ti}$, the tidal components of the velocity field, follow from solving the dominant momentum and mass balances for tidal hydrodynamics (see e.g. \citeA{ianniello1979}), which read
\begin{subequations} \label{eq:hyti}
	\begin{eqnarray}
		&& \frac{\partial u_{ti}}{\partial t} = - g \frac{\partial \eta_{ti}}{\partial x} + A_{v,ti} \frac{\partial^2 u_{ti}}{\partial z^2} ,\\
		&&\frac{1}{b} \frac{\partial }{\partial x} \left(b u_{ti}\right) + \frac{\partial w_{ti}}{\partial z} = 0 , \label{eq:tc}
	\end{eqnarray}
\end{subequations} 
in which $A_{v,ti}$ is the tidal vertical eddy viscosity. 
These equations are valid for externally forced tidal components. Note that the vertical viscosity coefficient differs from the one used for subtidal flow, which is appropriate for this kind of decomposition \cite{godin1991,godin1999}.
Boundary conditions are a harmonic water level at the estuary-sea transition, continuity of discharge and water level at the boundaries of the segments, no flow at the estuary-river transition (a weir is present in the cases we will consider), free slip at the surface, condition Eq.~\ref{eq:slip} for the bottom boundary (where $S_{f,st}$ is replaced with $S_{f,ti}$, the friction coefficient acting on the tidal flow, and $u_{st}$ with $u_{ti}$), and kinematic boundary conditions for $w_{ti}$. 
We refer again to \ref{app:vel} for specific formulation of boundary conditions and the analytical solutions of these equations. 

\subsubsection{Salinity}

Equations for the different components of the salinity in Eq.~\ref{eq:dec} are constructed from the width-averaged advection-diffusion equation for salinity, which reads 
\begin{align}\label{eq:sb} 
	\frac{\partial s}{\partial t} + \frac{1}{b} \frac{\partial}{\partial x} (b u s) + \frac{\partial}{\partial z} \Big( w s \Big) = \frac{1}{b} \frac{\partial}{\partial x} \Big(b K_h \frac{\partial s}{\partial x} \Big) + \frac{\partial}{\partial z} \Big( K_v \frac{\partial s}{\partial z} \Big).  
\end{align} 
Here, $K_h$ and $K_v$ are horizontal and vertical diffusivity coefficients, respectively. 
Vertical boundary conditions are no-flux at both the bottom $z=-H$ and the undisturbed water level $z=0$. At the river head, salinity is prescribed to have the river salinity. The condition at the mouth is derived in the same manner as in \citeA{biemond2022}, i.e. the estuarine domain is extended to include a simple representation of the adjacent sea. At the boundaries of the segments, continuity of salt transport and salinity, both for the tidal and subtidal components, is imposed. 

To obtain the subtidal quantities, we average Eq.~\ref{eq:sb} over the tidal timescale. An equation for $\bar s_{st}$ follows when taking an average over depth and application of the boundary conditions. For $s'_{st}$, the equation for $\bar s_{st}$ is subtracted from the tidally-averaged equation. 
For $s_{ti}$, we subtract the tidally-averaged equation from Eq.~\ref{eq:sb}. This equation is simplified following the approximation procedure in \citeA{wei2016}. The result is an equation for the dominant tidal constituent of salinity. Our equation differs from the one found in \citeA{wei2016} due to the presence of subtidal stratification. 
The equations for the different salinity components are given in \ref{app:sal}.

\subsubsection{Parametrizations of eddy viscosity and diffusivity}

To complete the model, closure relations for viscosity and diffusivity have to be specified. 
For vertical viscosity and diffusivity, we employ the parametrizations proposed in \citeA{MC04} and \citeA{ralston08hudson}:
\begin{equation} \label{eq:turb}
	A_{v,st} = c_{v,st} U_T H, \quad A_{v,ti} = c_{v,ti} U_T H, \quad 
	K_{v,st} = \frac{A_{v,st}}{Sc_{st}}, \quad K_{v,ti} = \frac{A_{v,ti}}{Sc_{ti}},
\end{equation}
in which $c_{v,st}$ and $c_{v,ti}$ are constants, $U_T$ is the scale of the amplitude of the tidal current, and $Sc_{st}$ and $Sc_{ti}$ are Schmidt numbers. For the considered estuaries, a constant value of $U_T =  1$ m s$^{-1}$ is used, which is justified a posteriori by the calculated tidal velocities in the model. This leaves $c_{v,st}$, $c_{v,ti}$, $Sc_{st}$ and $Sc_{ti}$ as unknown constants. 
Regarding $K_{h,st}$, we can not use closure relations like those in \citeA{banas04}, \citeA{MC07}, and \citeA{aristizabal15}, as these include all effects of tides, whilst our model resolves longitudinal and vertical tidal motion. However, dispersion can not be left out completely, as e.g. other tidal constituents and correlations between lateral variations of velocity and salinity are also represented by this process. Therefore, we use a simple constant value for $K_{h,st}$.  

\subsubsection{Solution procedure}

The solution procedure is as follows. 
First, using analytical expressions, the velocities are calculated (the exchange flow still being a function of subtidal salinity), and these are subsequently inserted into the equations for salinity. 
The equation for $s_{ti}$ (Eq.~\ref{eq:sti}) is a forced diffusion equation. Its solution $s_{ti}$ is obtained as a function of $s_{st}$, by standard techniques for non-homogeneous ordinary differential equations. The resulting expression for $s_{ti}$ is then substituted into the equations for $\bar s_{st}$ and $s'_{st}$ (Eqs.~\ref{eq:sstb} and \ref{eq:sstp}), which are subsequently, together with the boundary conditions, solved numerically for $s_{st}$ using the technique explained in \citeA{biemond2022}. The numerical discretization techniques are a Galerkin method in the vertical, a central difference scheme for the horizontal derivatives, and Crank-Nicolson \cite{crank1947} for time derivatives. 

\subsection{Field data}

\begin{figure}[]
	\centering
	\includegraphics[width=\textwidth]{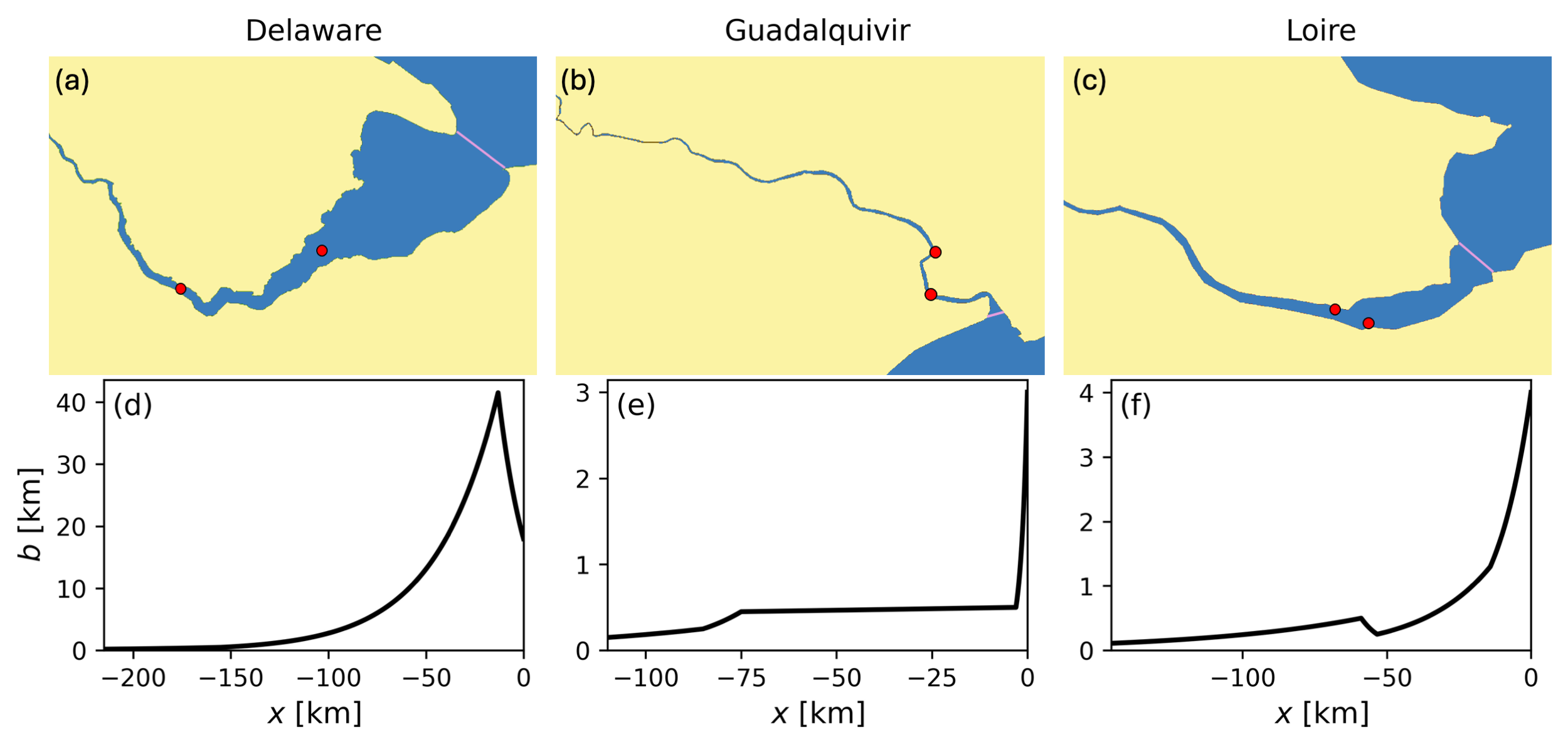}
	\caption{(a)-(c) Geometries of the Delaware, Guadalquivir and Loire estuaries, respectively. The pink lines indicate where the mouths of the estuary are defined in the model. (d) Width of the Delaware Estuary in the model versus $x$. (e)-(f) As (a), but for the Guadalquivir and Loire estuaries, respectively. 
		\label{fig:map}}
\end{figure}

For a model-data comparison (the first research aim), data of the geometry and forcing conditions (time series of river discharge and amplitude of the sea level at the mouth) of different estuaries are required. Observations of salinity and water levels were used to evaluate the model performance. 

Maps of the three selected estuaries (Delaware, Guadalquivir, Loire) are presented in Fig.~\ref{fig:map}a-c. The widths of the estuaries are estimated from satellite images and presented in Fig.~\ref{fig:map}d-f. A representative depth of the Delaware Estuary is 7.4 m and the Guadalquivir Estuary is 7.1 m deep. The Loire Estuary needs some specific attention, as longitudinal variations in its depth are substantial. Therefore, three different depths were used: the most downstream segment, which is 14 km long, is 12 m deep, the next 45 km long segment is 9 m deep, and the remainder of the estuary is 2.5 m deep \cite{walther2012}. 

For the Delaware Estuary, discharge at Belvidere, NJ, and salinity at three locations were extracted from the USGS database ( \url{https://waterdata.usgs.gov/nwis}). Results of the sea level tidal amplitudes and phases reconstructed from observations were taken from \citeA{fried94}. 
Regarding the Guadalquivir Estuary, we used the discharge at the Alcaca del Rio dam, as published by the `Agencia de Medio Ambiente y Agua de Andalucía'. Observed time series of subtidal salinity at seven observation points are provided by \citeA{navarro2011temporal,navarro2012use}. The tidal data (available at eight locations) are from \citeA{diez12}. 
In the Loire Estuary, discharge at Montjean-sur-Loire and water levels at five observation points upstream of Nantes were taken from \mbox{\url{hydro.eaufrance.fr}}. Water levels at six points downstream of Nantes are from \url{data.shom.fr}. The harmonic components of the $M_2$ tide were extracted from these data by the TTide software \cite{pawlowicz02}. Finally, time series of subtidal salinity salinity were provided by \citeA{Grasso23} (at five locations). 

For each of these estuaries, a dry summer was selected as calibration period. In the Delaware Estuary, the summer of 2023 was selected, the summer of 2009 for the Guadalquivir Estuary and 2013 for the Loire Estuary. During the calibration period in the Guadalquivir Estuary, water was extracted from the estuary for irrigation \cite{rep_gua_09}. We took this partly into account by subtracting 50 m$^3$s$^{-1}$ from the discharge in the period May 4-18, but this is probably an underestimation of the total water used for irrigation during this summer. 

%--------------------------
\section{Model Results} 
\label{sec:res}
%-----------------------

\subsection{Representation of tides and salinity}

\subsubsection{Method}

To demonstrate that the developed model is able to accurately represent hydrodynamic and salinity characteristics of different estuaries (the first research aim), the model is first calibrated for each estuary. The model is forced with observed time series of river discharge and tidal water level variations at the mouth. Salinity at the ocean boundary $s_{oc}$ is set to 35 psu, and salinity at the river boundary is 0.15 psu for the Delaware and Loire estuaries and 0.5 psu for the Guadalquivir Estuary. Regarding the numerical settings, the horizontal grid size varies per segment, and has a typical value of a few hundred meters, while in the vertical $z$-direction five Fourier modes are used, and the model timestep is one day. %The calibration procedure is detailed in Appendix C.
The value of $c_{v,ti}$ is determined by minimizing the root-mean-squared difference between the observed and modeled water level variations of the dominant tidal constituent, using a gradient descent algorithm. Likewise, the values of $c_{v,st}$, $K_{h,st}$, $Sc_{st}$ and $Sc_{ti}$ are found by minimizing the root-mean-squared difference between the observed and modeled subtidal salinity over a dry summer. 
For these simulations, the model is forced with the observed river discharge time series, and the observed sea level amplitude of the M$_2$ tide at the mouth, which is 0.68 m for the Delaware Estuary, 0.95 m for the Guadalquivir Estuary, and 1.80 m for the Loire Estuary. 
The model skill regarding salinity is then quantified by the refined index of agreement (IA) defined by \citeA{Willmott2012}. 
For this index, a value of 1 indicates perfect agreement and a value of -1 indicates that the model is poor.

\subsubsection{Result}

\begin{figure}[]
	\centering
	\includegraphics[width=\textwidth]{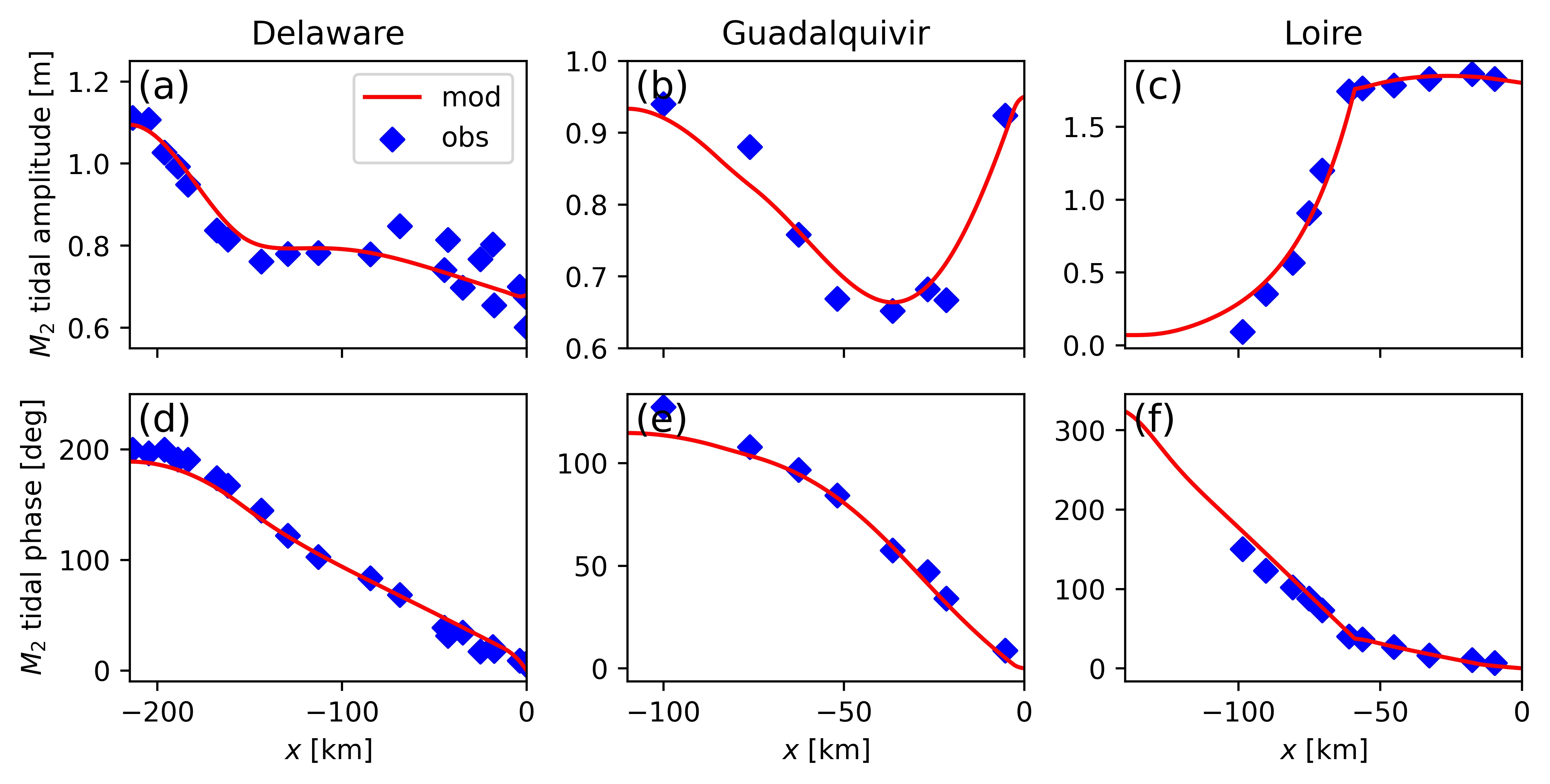}
	\caption{(a) Observed and modeled amplitude of the $M_2$ tidal water level in the Delaware Estuary as a function of of $x$, where the absolute value of $x$ is the distance to the mouth. The blue dots indicate the observations and the red line the modeled values. (b)-(c) As (a), but for the Guadalquivir and the Loire estuaries, respectively. (d)-(f) As (a)-(c), but for the tidal phase.
		\label{fig:cal_tid}}
\end{figure}

The observed and modeled characteristics of the M$_2$ component of the tidal water level in the Delaware, Guadalquivir and Loire estuaries are shown in Fig.~\ref{fig:cal_tid}, and the values for $c_{v,ti}$ are given in Table~\ref{tab:cal}. The main characteristics of the water level variations are resolved, i.e. the almost constant tidal strength in the downstream 150 km of the Delaware Estuary due to a balance between friction and channel convergence, reflection of the tidal wave at the weir at the upstream boundaries of the Delaware and Guadalquivir estuaries, and the dampening of the tide in the middle of the Guadalquivir Estuary and the upstream part of the Loire Estuary. The scatter in observed amplitude of the tide for the seaward side of Delaware estuary is because measurements are taken at both embankments and the Coriolis force affects the tides differently at the northern side of this wide estuary than at the southern side \cite{Lee2017}. There is also little difference regarding tidal phase speed between the model and the observations. These findings give confidence in the use of the modeled tidal hydrodynamics for salt intrusion modeling. 

\begin{table}[htbp]	
	\centering
	\caption{Values of the parameters of the turbulence closure for the different estuaries.}
	\label{tab:cal}
	\begin{tabular}{lccc}
		\hline
		Quantity & Delaware & Guadalquivir & Loire   \\
		\hline
		$c_{v,ti}$ & $1.3 \cdot 10^{-3}$ & $1.2 \cdot 10^{-3}$ & $2.3 \cdot 10^{-3}$  \\	
		$Sc_{ti}$ & 2.2 & 1.9 & 2.2  \\	
		%	$K_{h,ti}$ [m$^2$s$^{-1}$]& 20 & 20 & 20  \\
		$c_{v,st}$ & $5.5 \cdot 10^{-5}$& $1.0 \cdot 10^{-4}$ & $2.2 \cdot 10^{-4}$  \\	
		$Sc_{st}$ & 2.1 & 2.2 & 2.2  \\	
		$K_{h,st}$ [m$^2$s$^{-1}$]& 25 & 22 & 44  \\	
		\hline
	\end{tabular}
\end{table}

\begin{figure}[]
	\centering
	\includegraphics[width=\textwidth]{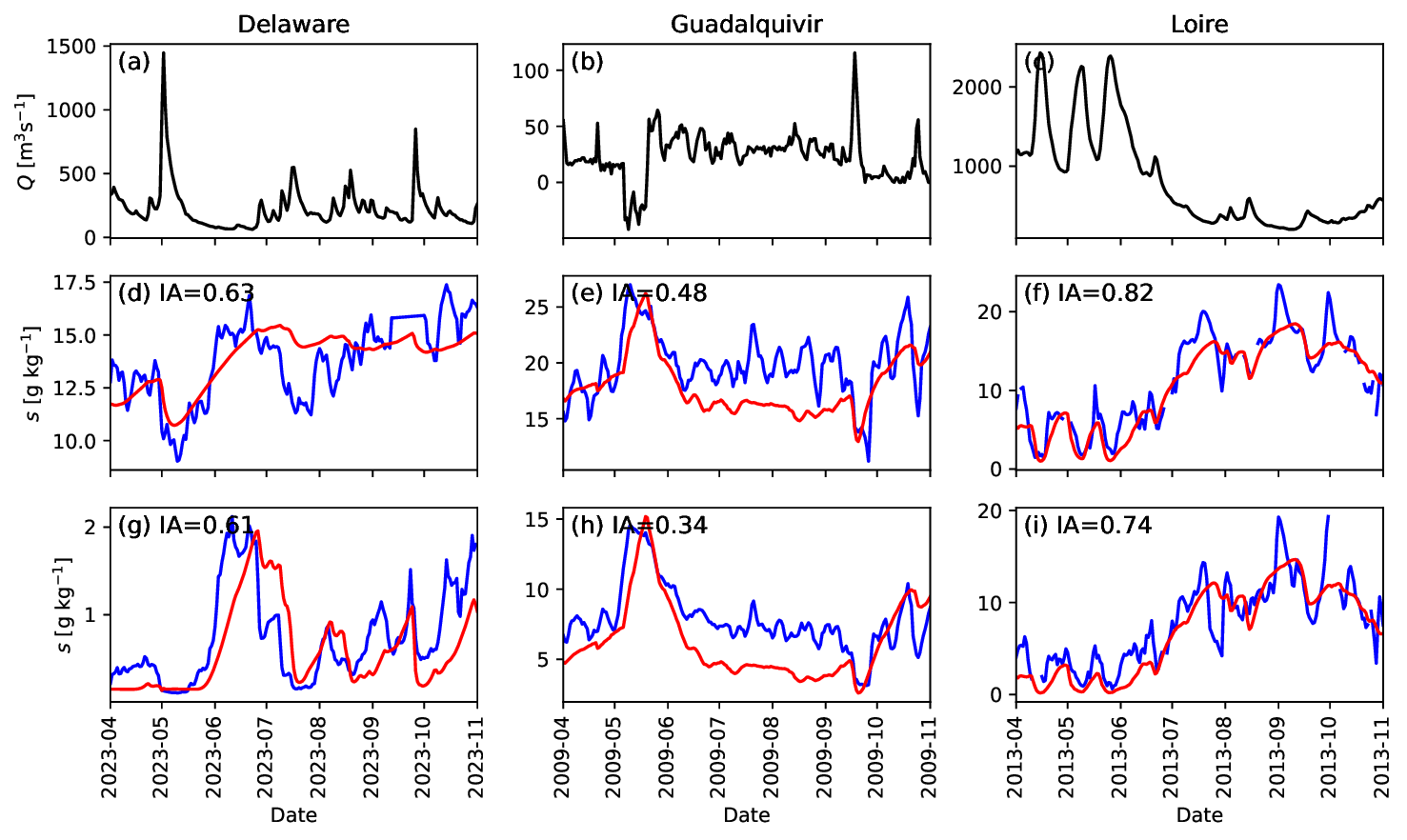}
	\caption{(a)-(c) Discharge time series for the Delaware, Guadalquivir and Loire estuaries, respectively. (d) Surface salinity time series from observations (in blue) and from the model (in red) for an observation station 62 km from the sea boundary in the Delaware Estuary. The refined index of agreement IA \cite{Willmott2012} is also indicated. (e) As (d), but for a station 17 km from the sea boundary in the Guadalquivir Estuary. (f) As (d), but for a station 12 km from the sea boundary of the Loire Estuary. (g)-(i) As (d)-(f), but for stations 110, 35 and 15 km from the sea boundaries, respectively. The locations of the shown salinity observation points are indicated with red dots in Fig.~\ref{fig:map}. 
		\label{fig:cal_sal}}
\end{figure}

%salinity
The values of the coefficients that determine vertical eddy viscosity and eddy diffusivity after calibration can be found in Table~\ref{tab:cal}.
Fig.~\ref{fig:cal_sal}a-c displays discharge during the calibration period for the three estuaries. In spring, freshwater discharge is higher than in summer in the Delaware and Loire estuaries. In the Guadalquivir Estuary, the freshwater intakes for irrigation early May are clearly visible as a drop in the discharge. 
Observed and modeled salinity are displayed in Fig.~\ref{fig:cal_sal}d-i. 
The model captures the variation in time of salinity in the Loire Estuary well ($IA\approx0.8$). The model skill is lower, but still reasonable, for the Delaware Estuary ($IA\approx0.6$) and for the Guadalquivir Estuary  ($IA\approx0.4$). 
This lower skill for the latter estuaries probably originates from missing forcing conditions, i.e. substantial differences are present between the applied discharge in the model and the discharge in reality.
For the Delaware Estuary, tributaries contribute to the total discharge \cite{aristizabal15} and are neglected in the model forcing.
In the Guadalquivir Estuary, water extractions for irrigation decrease the discharge, which leads to an underestimation of salinity during the dry period. 

%-----------------------
\subsection{Quantification of salt transports due to different mechanisms} \label{sec:saltrans}
%-----------------------

\subsubsection{Method}

The magnitudes of different salt salt transport components that contribute to the upstream salt transport (the second research aim) are calculated for equilibrium simulations using Eq.~\ref{eq:Tdec}. As forcing conditions we use the observed sea level amplitudes at the mouth, and river discharge that represents the dry season. We use for the Delaware Estuary $Q_{riv}~=~200$~m$^3$s$^{-1}$, for the Guadalquivir Estuary $Q_{riv}~=~15$~m$^3$s$^{-1}$ and for the Loire Estuary $Q_{riv}~=~350$~m$^3$s$^{-1}$. Note that the forcing conditions for the estuaries change through time and that the solutions from these equilibrium simulations will not be obtained most of the time in reality. However, inertia of the salt field complicates the analysis of the salt transport, and therefore equilibrium simulations are preferred for this research question. 

For quantification of the interaction between salt transports due to exchange flow and tidal flow, a series of additional simulations is performed without salt transport due to tidal advection, and a series without salt transport due to exchange flow. From the components of the salt transport in these simulations, two metrics are defined.
The effect of the tidal flow on the salt transport by the exchange flow is quantified with the use of the metric $I_{t, e}$, which reads
\begin{equation} \label{eq:ite}
	I_{t,e} = \frac{T_{e} - \tilde T_e}{\min{(T_e, \tilde T_e)}}.
\end{equation}
In this expression, $\tilde T_e$ is  the salt transport due to exchange flow in a simulation where tidal advection is set to zero. %A minimum is taken in the denominator, because both quantities are negative. 
A positive (negative) value of $I_{t,e}$ means that the presence of tides increases (decreases) the magnitude of $T_e$, i.e. it becomes more (less) negative.  
The effect of exchange flow on tidal salt transport is quantified in the same manner, with metric $I_{e,t}$ which reads
\begin{equation} \label{eq:iet}
	I_{e,t} = \frac{T_{\overline{ti}} -\tilde T_{\overline{ti}}}{\min{(T_{\overline{ti}} , \tilde T_{\overline{ti}})}},
\end{equation}
where $\tilde T_{\overline{ti}}$ is the salt transport due to time correlation of depth-averaged tidal flow and depth-averaged salinity in a model simulation without exchange flow. A positive (negative) value of $I_{e,t}$ means that the presence of exchange flow increases (decreases) the magnitude of $T_{\overline{ti}}$, i.e. it becomes more (less) negative.  
Note that we do not investigate the interactions between $T_e$ and $T_{ti'}$ or between $T_{\overline{ti}}$ and $T_{ti'}$, as $T_{ti'}$ turned out to be small with respect to $T_e$ and $T_{\overline{ti}}$ in the cases we consider. 

\begin{figure}[]
	\centering
	\includegraphics[width=\textwidth]{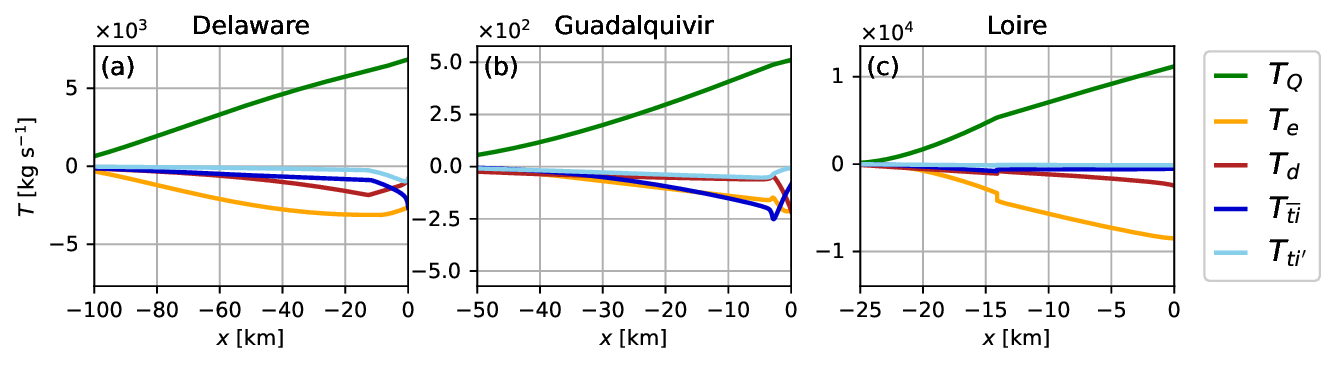}
	\caption{(a) The different components of the salt transport in equilibrium, following the decomposition of Eq.~\ref{eq:Tdec}, versus $x$ for the Delaware Estuary. (b)-(c) As (a), but for the Guadalquivir and Loire estuaries, respectively.  
		\label{fig:proc_all}}
\end{figure}

\subsubsection{Result}

The components of the equilibrium salt transport in the three considered estuaries are shown in Fig.~\ref{fig:proc_all}; the associated salinity structures $\bar s_{st}(x)$ are in Supporting Figure~S1.
The differences in the total downstream transport $T_Q$ between the estuaries originate from the differences in discharge. In the Delaware Estuary (panel a), upstream salt transport by the exchange flow ($T_e$) is dominant, while the other components are smaller. In the narrow part close to the sea boundary, the magnitudes of salt transport due to correlations in time between the depth-averaged tidal flow and salinity ($T_{\overline{ti}}$) and the salt transport due to correlations in time of the departures of the tidal flow and salinity from their depth-averaged values ($T_{ti'}$) increase. In the widest part of the estuary (around \mbox{$x=-30$ km}) salt transport by horizontal mixing $T_D$ is the second largest contributor to the upstream salt transport.
In the Guadalquivir Estuary (panel b), $T_{\overline{ti}}$ and $T_e$ are about equally strong in most of the estuary. At the change in width convergence around \mbox{$x=-3$ km}, $T_{\overline{ti}}$ has a maximum. The other components ($T_d$ and $T_{ti'}$) are negligible, except close to the sea boundary, where the magnitude of $T_d$ increases strongly. 
Panel c shows that, for the Loire Estuary, $T_e$ is very dominant compared to the other salt transport components. Despite the strong tidal water level variations, the tidal salt transport is thus negligible here.

\begin{figure}[]
	\centering
	\includegraphics[width=\textwidth]{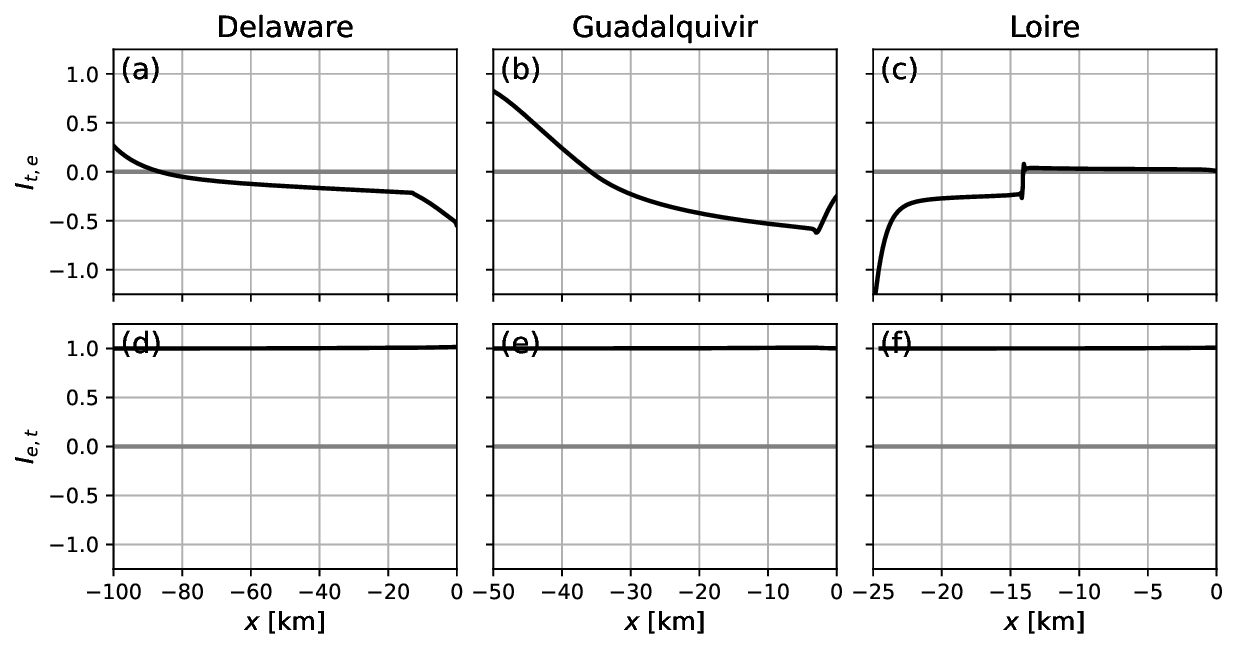}
	\caption{(a) Effect of tidal advection on the salt transport by exchange flow, quantified by $I_{t,e}$ (Eq.~\ref{eq:ite}), versus $x$ for the Delaware Estuary. (b)-(c) As (a), but for the Guadalquivir and Loire estuaries, respectively. (d)-(f) As (a)-(c), but for the effect of exchange flow on salt transport by tidal flow, quantified with $I_{e,t}$ (Eq.~\ref{eq:iet}). % For $\bar s_{st}<0.33$, the values are not calculated, which is visible for the Loire Estuary upstream of $x=-20$ km. 
		\label{fig:int}}
\end{figure}

The indices that measure the interaction of $T_e$ and $T_{\overline{ti}}$ (see Eqs.~\ref{eq:ite} and \ref{eq:iet}) are shown in Fig.~\ref{fig:int} and the values of $\tilde T_e$ and $\tilde T_{\overline{ti}}$ can be found in Supporting Figure S2. Note that the extent of salt intrusion changes substantially when excluding salt transport components (Fig.~S1). Figs.~\ref{fig:int}a-b show that the presence of tidal salt transport decreases the magnitude of $T_e$ downstream in the Delaware and Guadalquivir estuaries, but increases it further upstream. In the Loire Estuary (panel c), the effect of tidal salt transport is small downstream of $x=-14$ km, where a change in depth is located, and increases upstream of this point. 
The values of $I_{e,t}$ are close to one (panels d-f), i.e. variations are below 0.01, much smaller than the variations in $I_{t,e}$. This indicates that without exchange flow $T_{\overline{ti}}$ is absent. These results will be explained in Section~\ref{sec:mag}. 

%-----------------------
\section{Interpretation of the model results} 
\label{sec:resi}
%-----------------------

\subsection{Explanation of the magnitude of the salt transport components}
\label{sec:mag}
%-----------------------

\subsubsection{Method}

To gain more insight into the magnitudes of the different salt transport components (the third research aim), we analyze the characteristics of the two largest upstream salt transport components in Fig.~\ref{fig:proc_all}, i.e. $T_e$ and $T_{\overline{ti}}$, making use of approximate analytical expressions for these components. 
To construct those, we first find expressions for $s'_{st}$ and $s_{ti}$, which follow when assuming that subtidal stratification is the result of a balance between advection of $\bar s_{st}$ by the vertical shear of the subtidal current and vertical diffusion. This balance was found by \citeA{pritchard1954} from analysis of field data, and hereafter we will refer it as the \citeA{pritchard1954} approximation. The resulting expressions for $s'_{st}$ and $s_{ti}$ are presented in \ref{app:pri}. A confirmation that the salt transport components are well approximated under this assumption follows from comparing Fig.~\ref{fig:proc_all} with Supporting Figure S2g-i. 

Using the \citeA{pritchard1954} approximation, it follows \cite{MC04}
\begin{equation} \label{eq:te}
	T_e = b H \overline{u'_{st} s'_{st}} \approx - \frac{\alpha^2 b H^3 }{K_{v,st}} 0.112 \left(\frac{\partial \bar s_{st}}{\partial x}\right)^3  ,
\end{equation}
for the salt transport due to exchange flow, where $\alpha = \frac{g \beta H^3}{48 A_{v,st}} $. This quantity thus scales with width, horizontal salinity gradient, stratification (whose variation is contained in the horizontal salinity gradient) and depth. 
%
%derivation expression tidal velocity
The salt transport component $T_{\overline{ti}}$ reads
\begin{equation} \label{eq:tti1}
	T_{\overline{ti}} = b H (\bar{u}_{ti} \bar{s}_{ti})_{st} = \frac{bH}{2} |\hat{\bar{u}}_{ti}| |\hat{\bar{s}}_{ti}| \cos(\Delta \theta_{us}), 
\end{equation}
in which $|\hat{\bar{u}}_{ti}|$ and $|\hat{\bar{s}}_{ti}|$ are the amplitudes of the tidal variations of depth-averaged flow and salinity, respectively, and $\Delta \theta_{us}$ is the phase difference between depth-averaged tidal flow and salinity. 
Note that $T_{\overline{ti}}$ will be directed upstream, which is the case in our model results, if $\Delta \theta_{us}$ is larger than $90^{\circ}$. 
To derive a more easily interpretable expression, we first find an equation for $\bar s_{ti}$ by taking a depth-average of Eq.~\ref{eq:sti}:
\begin{equation} \label{eq:bsti}
	\frac{\partial \bar s_{ti}  }{\partial t} + \bar u_{ti} \frac{\partial \bar s_{st}}{\partial  x}  + \overline{w_{ti} \frac{\partial  s'_{st}}{\partial  z} }  =  0.
\end{equation}
To find an approximation for the amplitude of $\bar s_{ti}$, we note that the vertical advection term is much smaller than the horizontal advection term. Thus, to a first approximation, the amplitude $\hat{\bar s}_{ti}$ of $\bar s_{ti}$ reads $\frac{1}{\omega} |\bar u_{ti}| \frac{\partial \bar s_{st}}{\partial x}$. In this expression, $\omega$ is the angular frequency of the tidal constituent.
Substituting this in Eq.~\ref{eq:tti1} yields
\begin{equation} \label{eq:TTb}
	T_{\overline{ti}} \approx \frac{b H}{2 \omega} |\hat{\bar u}_{ti}|^2 \frac{\partial \bar s_{st}}{\partial x} \cos(\Delta \theta_{us}). 
\end{equation}
An alternative expression for $T_{\overline{ti}}$ is derived in \ref{app:Tti} and reads 
\begin{equation} \label{eq:tti}
	%	T_{\overline{ti}} \approx \frac{b}{2} |\hat{\bar u}_{ti}| \ |\hat \eta_{ti}| \ |s'_{st} (z=0)| \cos \left(\arg(\hat \eta_{ti}) - \arg(\hat{\bar u}_{ti})\right).
	T_{\overline{ti}} \approx \frac{23}{300} \frac{\alpha b H^2}{K_{v,st}} |\hat{\bar u}_{ti}| \ |\hat \eta_{ti}| \ (\frac{\partial \bar s_{st}}{\partial x})^2  \cos\left(\Delta \theta_{u\eta}\right) ,
\end{equation}
where $ |\hat \eta_{ti}|$ is the amplitude of the tidal water level and $\Delta \theta_{u\eta}$ is the phase difference between depth-averaged tidal current and tidal water level. The phase difference $\Delta \theta_{u\eta}$ is always more than 90 degrees (a property of a progressive tidal wave in the negative $x$-direction), which means that $T_{\overline{ti}}$ in Eq.~\ref{eq:tti} is negative, i.e. the transport is upstream. 
The ratio between $T_e$ and $T_{\overline{ti}}$ reads 
\begin{equation} \label{eq:rat}
	\frac{T_e}{T_{\overline{ti}}} = -1.46 \frac{\alpha H \frac{\partial \bar s_{st}}{\partial x}}{|\hat{\bar u}_{ti}| \ |\hat \eta_{ti}| \ \cos\left(\Delta \theta_{u\eta}\right) } ,
\end{equation}
which follows from dividing Eq.~\ref{eq:te} by Eq.~\ref{eq:tti}.

\subsubsection{Result}

\begin{figure}[]
	\centering
	\includegraphics[width=\textwidth]{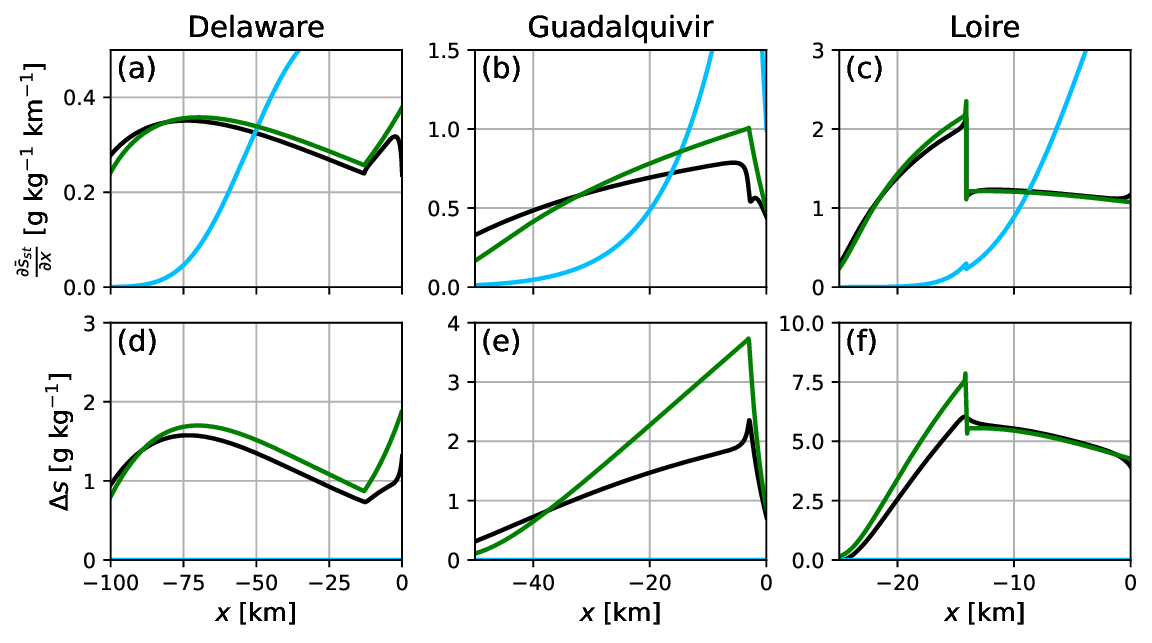}
	\caption{(a) Horizontal subtidal salinity gradient versus $x$ in the Delaware Estuary for different simulations. The black line indicates the simulation with the complete model, the blue line uses a version without stratification, and the green line uses a version without advective salt transport by tides. (b)-(c) As (a), but for the Guadalquivir and Loire estuaries, respectively. (d)-(f) As (a)-(c), but for subtidal stratification, i.e. the bottom-surface subtidal salinity difference. 
		\label{fig:dsd}}
\end{figure}

The longitudinal variation of $T_e$ is completely determined by the variation in horizontal salinity gradient $\frac{\partial \bar s_{st}}{\partial x}$ in an estuary with a constant cross-section (Eq.~\ref{eq:te}). We present the horizontal salinity gradient $\frac{\partial \bar s_{st}}{\partial x}$ for the three estuaries in Fig.~\ref{fig:dsd}a-c. 
A comparison between $\frac{\partial \bar s_{st}}{\partial x}$ and $T_e$ (Fig.~\ref{fig:proc_all}) shows that the variation in $\frac{\partial \bar s_{st}}{\partial x}$ indeed dominates the variation in $T_e$ in the Delaware and Guadalquivir estuaries. However, far downstream in these estuaries, and in the Loire Estuary, $T_e$ does not follow the variation of $\frac{\partial \bar s_{st}}{\partial x}$. To explain this, consider e.g. the Loire Estuary, where the longitudinal variation of $T_e$ is large compared to the variation of $\frac{\partial \bar s_{st}}{\partial x}$ in the downstream 14 km. The increase of the magnitude of $T_e$ with increasing $x$ is instead accompanied by an increase in width (Fig.~\ref{fig:map}f). Similar reasoning applies to the downstream parts of the Delaware and Guadalquivir estuaries.  

\begin{figure}[]
	\centering
	\includegraphics[width=\textwidth]{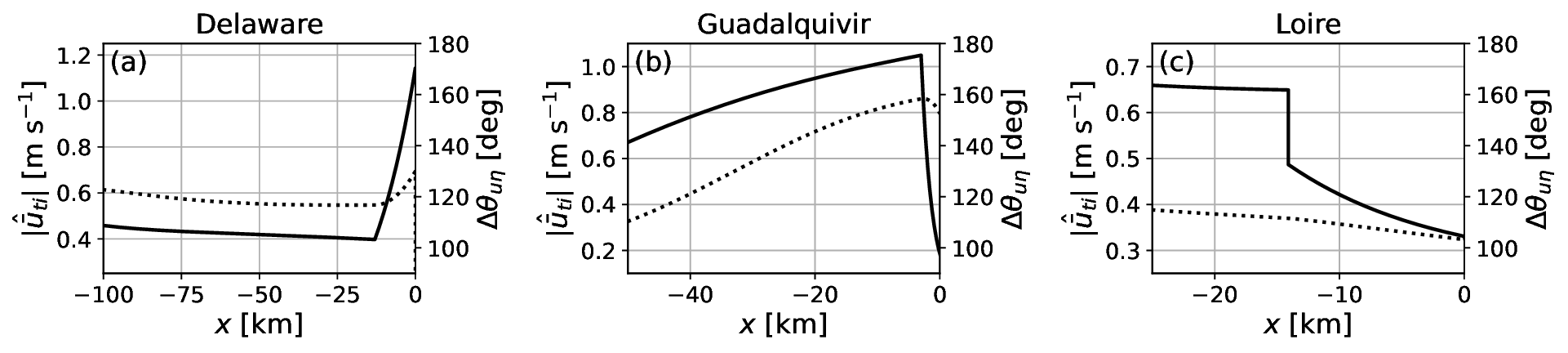}
	\caption{(a) Depth-averaged tidal current amplitude $|\hat{\bar u}_{ti}|$ (solid line) and its phase difference with the tidal water level $\Delta \theta_{u\eta}$ (dotted line) as a function of $x$, for the Delaware Estuary. (b)-(c) As (a), but for the Guadalquivir and Loire estuaries, respectively. 
		\label{fig:uuu}}
\end{figure}

Fig.~\ref{fig:proc_all}a shows that in the Delaware Estuary, the magnitude of $T_{\overline{ti}}$ increases with increasing $x$. To explain this, we use Eq.~\ref{eq:tti}. The relevant properties of the tidal currents and water levels are shown in Fig.~\ref{fig:uuu}, whose variations with $x$ are smaller than the variation of $T_{\overline{ti}}$, except close to the downstream boundary, where an increase in $|\hat{\bar u}_{ti}|$ is associated with an increase in the magnitude of $T_{\overline{ti}}$. In the rest of the domain, the variation of $T_{\overline{ti}}$ is best explained by the variation in width (Fig.~\ref{fig:map}a). 
A similar explanation holds for the Guadalquivir Estuary, but here the increase of the magnitude of $T_{\overline{ti}}$ with increasing $x$ is additionally strengthened by an increase in $|\hat{\bar u}_{ti}|$ (Fig.~\ref{fig:uuu}b). Close to the downstream boundary, the situation is reversed with respect to the Delaware Estuary: $|\hat{\bar u}_{ti}|$ has a minimum, associated with a decrease in the magnitude of $T_{\overline{ti}}$. 
In the Loire Estuary, $T_{\overline{ti}}$ is relatively weak with respect to $T_e$ (Fig.~\ref{fig:proc_all}c). 
Eq.~\ref{eq:rat} shows that the ratio $\frac{T_e}{T_{\overline{ti}}}$ increases with deeper estuaries and estuaries with a stronger salinity gradient. On the other hand, it decreases with magnitude of the tidal currents and magnitude of the variations in water level. Especially the effect of depth is large. 
The Loire Estuary is the deepest of the considered estuaries and has the strongest horizontal salinity gradient (Fig.~\ref{fig:dsd}c), which explains why $\frac{T_e}{T_{\overline{ti}}}$ is large in this estuary. 
On the other hand, in the Guadalquivir Estuary this ratio is relatively small, since this is the shallowest estuary and tidal currents are relatively strong (Fig.~\ref{fig:uuu}b).

To explain the interaction between the different components of salt transport, we first focus on the effects of tidal advection on $T_e$.  For this, we analyze the effects of tidal advection on horizontal salinity gradient $\frac{\partial \bar s_{st}}{\partial x}$ and stratification, which in turn affect $T_e$. 
The magnitudes of these quantities are larger in the downstream reach of the Delaware and Guadalquivir estuaries in the simulation without tidal advection, compared to the simulation including tidal advection (Fig.~\ref{fig:dsd}). However, further upstream, they are smaller in the simulation without tides. This is similar to the longitudinal variation of $I_{t,e}$ (Fig.~\ref{fig:int}a-b). 
 In the Loire Estuary, the difference between the two simulations is small, because $T_{\overline{ti}}$ is weak with respect to $T_e$. %The maximum changes are in order of 20\% for horizontal gradients and 50\% for stratification.
The changes in the $\frac{\partial \bar s_{st}}{\partial x}$ quantitatively explain the difference between the two simulations. For example, in the Guadalquivir Estuary around $x=-3$ km, $\frac{\partial \bar s_{st}}{\partial x}$ is reduced by 20\% when including tidal advection (Fig.~\ref{fig:dsd}b). According to Eq.~\ref{eq:te}, $T_e$ scales with $(\frac{\partial \bar s_{st}}{\partial x})^3$, which implies that a 20\% reduction of $\frac{\partial \bar s_{st}}{\partial x}$ reduces $T_e$ by about 50\%, which is indeed visible in Fig.~\ref{fig:int}b.
We thus find that, as expected, tidal advection reduces the horizontal salinity gradient, which weakens stratification and this leads to a decrease in $T_e$. Other, more direct effects of tidal advection on stratification are shown to be less important (Supporting Information S1).

\begin{figure}[]
	\centering
	\includegraphics[width=0.8\textwidth]{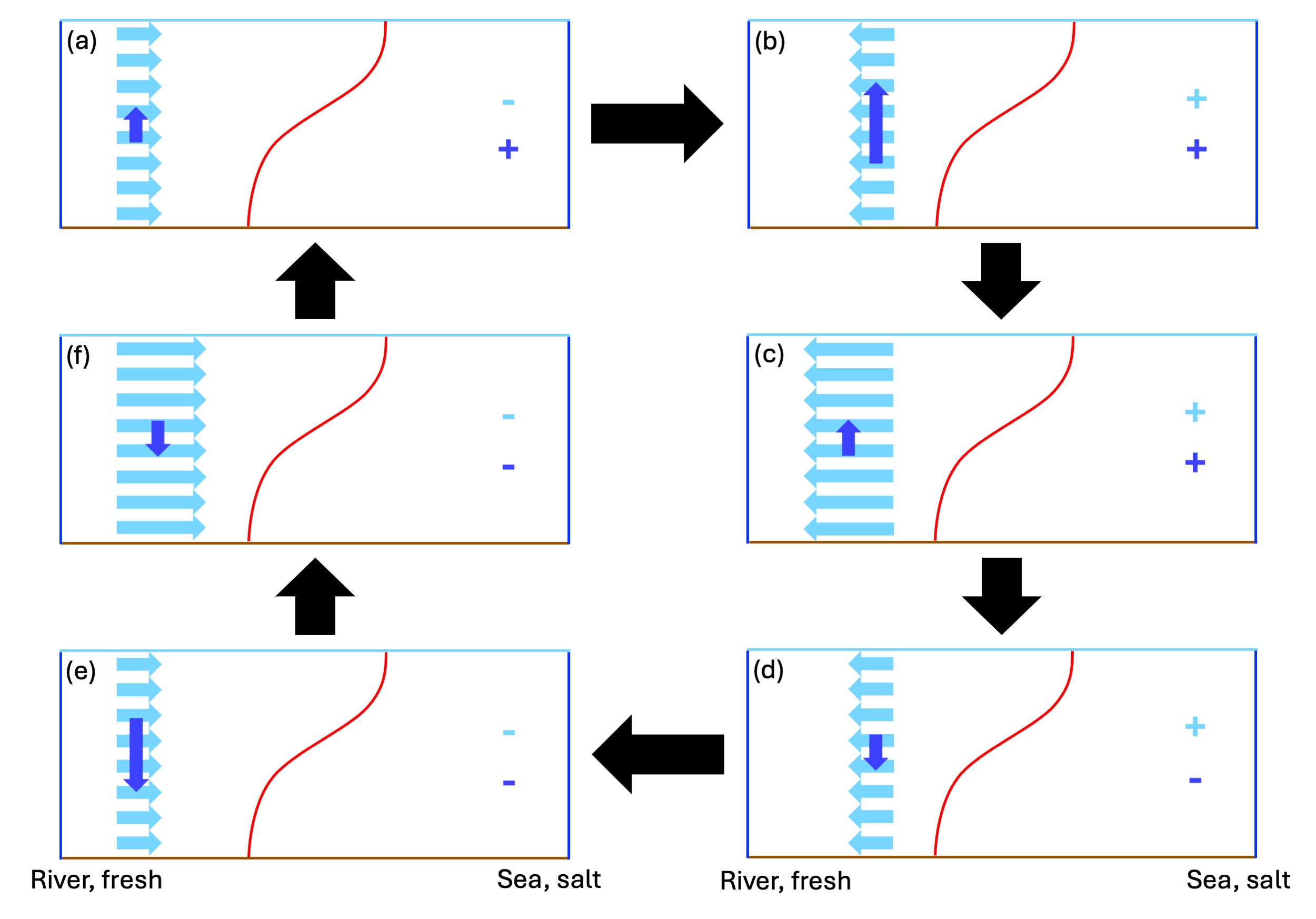}
	\caption{ Sketch of the dynamics represented by Eq.~\ref{eq:bsti}. The domain is a longitudinal section of an estuary. The light blue arrows indicate the depth-averaged horizontal tidal current, the dark blue arrows indicate the vertical component of the tidal current, and the red line is a subtidal isohaline. The light blue plus and minus signs indicate if the depth-averaged horizontal tidal current contributes to a decrease or an increase of the depth-averaged tidally varying salinity, respectively. The dark blue plus and minus signs are similarly defined, but for the vertical component of the tidal current. The different panels show different phases of the tide, i.e. (a) late flood, (b) early ebb, (c) middle ebb, (d) late ebb, (e) early flood, and (f) middle flood.
		\label{fig:concept}}
\end{figure}

The effect of exchange flow on $T_{\overline{ti}}$ is explained by considering Eq.~\ref{eq:bsti}, which describes the behavior of $\bar s_{ti}$, subject to tidal currents and subtidal salinity gradients. 
Without vertical salinity gradients (a well-mixed estuary), the last term on the left hand side of this equation is zero. In that case, the solution $\bar s_{ti}$ is $90^{\circ}$ out of phase with $\bar u_{ti}$, resulting in $T_{\overline{ti}} = 0$ (see Eq.~\ref{eq:tti1}), as in the models of \citeA{mccarthy1993residual} and \citeA{wei2016}. Fig.~\ref{fig:proc_all} shows that in our model $T_{\overline{ti}}<0$, which means that there are phase differences between the tidal flow and salinity, which must originate from the vertical advection term. This term is linearly related to the subtidal stratification, highlighting the importance of stratification for the magnitude of $T_{\overline{ti}}$. 
Since the dominant driver for subtidal stratification is the exchange flow (see Fig.~\ref{fig:dsd}d-f), the magnitude of $T_{\overline{ti}}$ is to a large extent determined by the magnitude of  $T_e$.
Fig.~\ref{fig:concept} illustrates the underlying concept. When only the term $-\bar u_{ti} \frac{\partial  \bar s_{st}}{\partial  x}$ would be taken into account, salinity always decreases during ebb and increases during flood (the light blue elements in the figure). The vertical advection term $-\overline{w_{ti} \frac{\partial s'_{st}}{\partial  z} }$ increases (decreases) the depth-averaged tidal salinity when $w_{ti}$ is positive (negative). For a tidal wave with $\Delta \theta_{u\eta}> 90^\circ$, this contribution is out of phase with $\bar u_{ti}$, as expected, and makes that during ebb salinity is lower and during flood salinity is higher, which results in a net landward transport of salt.

%-----------------------
\subsection{Magnitudes of salt transport components from length scales}
\label{sec:ls}
%-----------------------

\subsubsection{Method}

The fourth research aim is addressed by the development of a method which allows for obtaining a first-order estimate of the importance of salt transport due to exchange flow, tidal flow, and horizontal diffusion. For this estimate, only primary quantities (i.e. geometry, forcing conditions and eddy viscosity and diffusivity coefficients, which follow from the water depth and amplitude of the tidal current (Eq.~\ref{eq:turb})) and no numerical methods are required. 
The procedure is as follows. An approximation of Eq.~\ref{eq:Tdec} is derived, which only contains $\bar s_{st}$ and model parameters, without making assumptions about the magnitude of the different components. For this, we use analytical expressions for the different components of $u$ as presented in \ref{app:vel}. 
Expressions for $s'_{st}$, $\bar s_{ti}$ and $s'_{ti}$ as a function of $\frac{\partial \bar s_{st}}{\partial x}$ follow from solutions to Eqs.~\ref{eq:sstb} and \ref{eq:sstp} using the \citeA{pritchard1954} approximation and are given in \ref{app:pri}. Subsequently, when defining dimensionless salinity $\Sigma = \frac{s_{st}}{s_{oc}}$ and a dimensionless coordinate $\lambda = L_{sc} x$, where $L_{sc}$ is the length scale of the salt intrusion, and assuming equilibrium and $T=0$, Eq.~\ref{eq:Tdec} reads
\begin{eqnarray} \label{eq:L}
	\Sigma = \left(\frac{L_{E3}}{L_{sc}}\right)^3 \left( \frac{\partial \Sigma}{\partial \lambda} \right)^3 &+& \left(  \left(\frac{L_{E2}}{L_{sc}}\right)^2 + \left(\frac{L_{T2}}{L_{sc}}\right)^2 + \left(\frac{L_{T4}}{L_{sc}}\right)^2 \right) \left( \frac{\partial \Sigma}{\partial \lambda} \right)^2 \nonumber \\ &+& \left(\frac{L_{E1}}{L_{sc}} + \frac{L_{T1}}{L_{sc}} + \frac{L_{T3}}{L_{sc}} + \frac{L_D}{L_{sc}}\right) \left( \frac{\partial \Sigma}{\partial \lambda} \right) .
\end{eqnarray}
The expressions for the length scales are given in \ref{app:ls}. This equation extends the one presented by \citeA{MC04}, his Eq.~15, in the sense that it contains additional tidal terms. The interpretation of the first four length scales in Eq.~\ref{eq:L} is in that study. In short, $L_D$ is the length scale that the salt intrusion would have when horizontal diffusion would be the only upstream salt transport mechanism. In \citeA{MC04}, all the tidal effects are contained in this term, while in our model it represents only the effects that are not captured by the advective terms.
Likewise, $L_{E1}$, $L_{E2}$ and $L_{E3}$ are the length scales of the salt intrusion due to exchange flow related to vertical shear of the river flow, joint action of the shear of the river flow and density-driven flow, and density-driven flow, respectively. The new length scales $L_{T1}$ to $L_{T4}$ measure the effects of salt intrusion due to tidal advection. 
The first two are the length scales that quantify the salt intrusion due to $T_{\overline{ti}}$ (depth-mean tidal current), and the third and fourth are associated with $T_{ti'}$ (shear in tidal current). The distinction between $L_{T1}$ and $L_{T2}$ is that $L_{T1}$ is related to a phase difference between salinity and the tidal current that originates from vertical advection of subtidal stratification induced by shear of the river flow, while for $L_{T2}$ this phase difference is caused by the density-driven flow. The same distinction is between $L_{T3}$ and $L_{T4}$. 
To determine $L_{sc}$, we use the fact that the values of $\Sigma$ and $\lambda$ in Eq.~\ref{eq:L} are of order 1. Substituting this in the equation, gives a cubic equation for $L_{sc}$.% for every $x$. 

\subsubsection{Result}

\begin{figure}[]
	\centering
	\includegraphics[width=\textwidth]{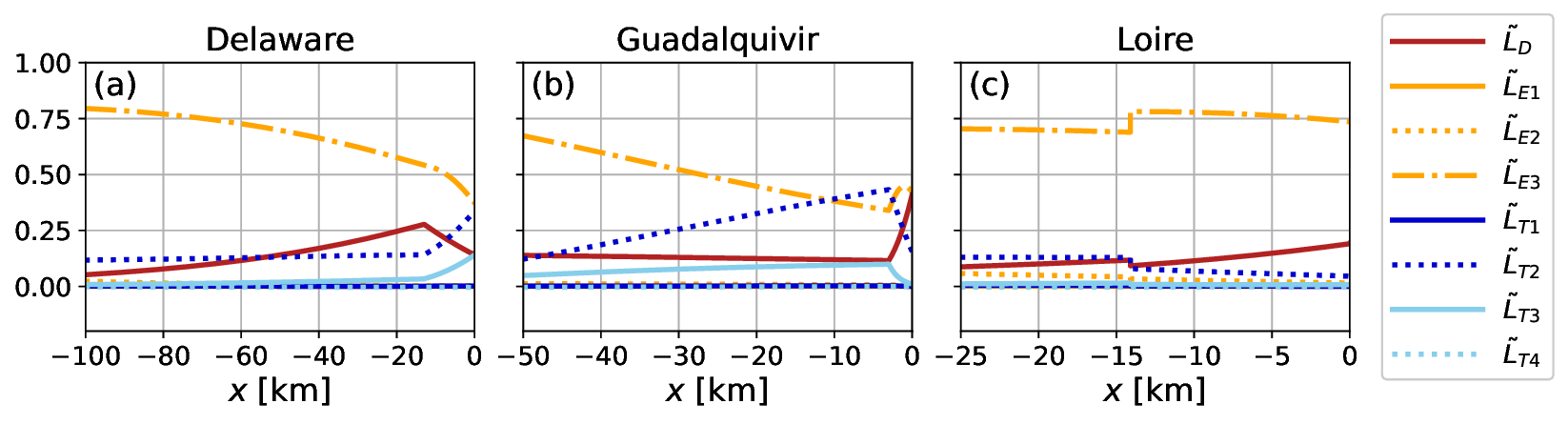}
	\caption{(a) The different terms in Eq.~\ref{eq:L} versus $x$ for the Delaware Estuary. Note that the normalized terms are plotted, e.g. label $\tilde L_D$ indicates $\frac{L_D}{L_{sc}}$. (b)-(c) As (a), but for the Guadalquivir and Loire estuaries, respectively. 
		\label{fig:ls}}
\end{figure}

The dimensionless length scale ratios in Eq.~\ref{eq:L} for the three estuaries are shown in Fig.~\ref{fig:ls}. %, where for $L_{sc}$ the length of the salt intrusion, i.e. the distance of the bottom 2 g kg$^{-1}$ isohaline to the estuary mouth, is used. 
The ratios of the different subtidal salt transport processes, as shown in Fig.~\ref{fig:proc_all}, are in line with what is shown in this figure: in the Delaware and Loire estuaries, salt transport due to exchange flow is dominant. In the Guadalquivir Estuary tidal salt transport is dominant from $x=-20$ to $x=-5$ km (the extent of this regime is a bit underestimated in Fig.~\ref{fig:ls}b), while close to the mouth and further upstream salt transport associated with the exchange flow is dominant. This agreement gives confidence in the use of this equation to analytically determine the importance of different salt transport processes. 
Regarding the terms associated with tidal salt transport in Eq \ref{eq:L}, $(\frac{L_{T2}}{L_{sc}})^2$ is the largest one. This indicates that stratification induced by the density-driven flow induces a larger distortion in the phase of the tidal salinity than the shear of the river flow, and that $T_{\overline{ti}} > T_{ti'}$, consistent with what is shown in Fig.~\ref{fig:proc_all}. 

%-----------------------
\section{Discussion} 
\label{sec:dis}
%-----------------------

We developed a semi-analytical model, which resolves salt transport by the exchange flow and tidal flow. Simpler models, like those developed by \citeA{MC04} and \citeA{wei2016} can not describe the interaction of those two types of salt transport. This interaction, however, is found to be important for the salt transport by the exchange flow as well as to the transport by the tidal flow (Section~\ref{sec:mag}). 
More complex models, such as the intermediate complexity model presented by \citeA{DSK22} and full-blown primitive equation 3D numerical models \cite{maccready2010,maccreadybanas11,chen12,gong2014,aristizabal2013,maccready24} do resolve the interaction of the different types of salt transport. However, their complexity makes it hard to find clear relationships between input and output variables. Also, deriving analytical expressions from the model equations for transports and their relative importance, as is done in this study (Sections \ref{sec:mag} and \ref{sec:ls}), is not straightforward in these models. 
Moreover, numerical diffusion can be substantial in a complex 3D model, which impacts the vertical salinity structure and therefore salt transport \cite{ralston17,wang21mix,schlichting23}.
The importance of numerical diffusion in our model is quantified by performing the simulations described in Section~\ref{sec:saltrans} with doubled vertical resolution, and comparing the subtidal salinity in these simulations with the original simulations. This gives a maximal difference of 0.5\% in the three estuaries considered, implying that numerical diffusion plays a negligible role for the salt transports calculated with our model.

Available observations of the relative importance of the components of the salt transport in the Delaware and Guadalquivir estuaries are mostly in agreement with the results from our modelling study. \citeA{aristizabal15} find, supported by detailed numerical modelling presented in \citeA{aristizabal2013}, that exchange flow is the dominant driver of salt import in this estuary, in line with our results (Fig.~\ref{fig:proc_all}a). Additionally, lateral shear of the currents also leads to substantial salt transport, which is in our model represented by $T_d$, and indeed substantially contributes to the upstream salt transport in the Delaware Estuary. 
\citeA{diez2013spa} observed that $T_{\overline{ti}}$ is dominant in the Guadalquivir Estuary. In our model (Fig.~\ref{fig:proc_all}b), this is the case in part of the estuary, but $T_e$ also contributes substantially, which is not the case in the observations. 
The reason for this difference may be that, on one hand, in our study only density-driven flow (and a minor contribution from the shear of the river flow) contribute to the estuarine exchange flow and thus to $T_e$, while other studies have shown that other processes like tidal straining also contribute to $T_e$ \cite{burchard10,burchard11}. Therefore, when these processes are present in reality, our model is likely to overestimate $T_e$. On the other hand, $T_{\overline{ti}}$ could be underestimated, because we only consider the dominant tidal constituent, while \citeA{diez2013spa} shows that other constituents also contribute to the salinity variation, which potentially increases the upstream tidal salt transport.

The analytical expressions for the magnitude of the different salt transport components that we derived highlight the importance of stratification for $T_{\overline{ti}}$. This relationship was also reported by \citeA{aristizabal2013}.
Other studies have also hold other factors, often geometrical effects, responsible for the phase shift between tidal currents and tidal salinity \cite{okubo1973,geyersignell92,macvean11}, which is not considered in our model. 
Moreover, the moving plane analysis introduced by \citeA{Dronkers1986} and further developed by \citeA{Garcia2023} shows that any process that affects salinity during its tidal excursion causes a phase difference and therefore salt transport by tidal currents.
Quantifying the role of different contributions to this phase difference is crucial to understand the salt transport associated with the tidal flow, and future work could extend the model developed in this study by taking the impact of geometrical features on the phases of tidal flow and salinity into account. 

The assumptions made to simplify the model equations allow to construct a fast and understandable model, but also limit its scope. An example where the model outcome will deviate substantially from reality is an estuary where Stokes transport of salt is important, i.e. the assumption regarding water level variations being much smaller than the mean water depth is violated. Other examples include salt wedge estuaries, where stratification is strong, and advection of momentum is important in the momentum balance \cite{DS21}, which is not taken into account in our model. 
Another limitation is that the model is forced with only river discharge and one tidal constituent. Other drivers of subtidal salt transport in estuaries exist, like storm surges \cite{yang2023}, and more tidal constituents could be required to describe the hydrodynamics and the associated salt transport to a reasonable extent \cite{li2023}. 
Finally, the turbulence closure chosen in this study is rather simple. In reality, the strength of the turbulence will depend on velocity shear and stratification, and therefore on location and time in the estuary. Estuaries where these variations play an important role for the salt transport are beyond the scope of this study, and this would be an interesting topic for future research. %can therefore not be described by the model. %Modelling these turbulence characteristics in detail requires a sophisticated formulation of viscosity and diffusivity, which will go at the cost of the understandability of the model. 

\section{Summary and conclusions}
\label{sec:con}

We constructed a width-averaged, semi-analytical model  for salt intrusion in estuaries, containing advective salt transport due to both exchange flow and tidal flow. With this model, we were able to hindcast observed tidal and salinity characteristics in three estuaries, i.e. the Delaware, Guadalquivir and Loire estuaries. We showed that, for the estuaries we considered, the salt transport due to the time correlations between vertical shear of the tidal current and vertical variation of tidal salinity is small compared to other salt transport components. The salt transport due to time correlation between the depth-averaged tidal current and the tidally varying depth-averaged salinity interacts strongly with salt transport due to the exchange flow. This interaction occurs through the stratification: the exchange flow generates stratification, which causes a phase difference between the tidal flow and the tidally varying part of the salinity, which results in a net upstream salt transport. The tidal flow, in turn, decreases the horizontal salinity gradient, which weakens the exchange flow. The magnitude of salt transport due to time correlation between the depth-averaged tidal current and the tidally varying depth-averaged salinity, compared to salt transport by the exchange flow, decreases with deeper estuaries and estuaries with a stronger salinity gradient. On the other hand, it increases with magnitude of the tidal currents and magnitude of the variations in water level. 
Using approximate analytical expressions for the components of the salt transport, we explained the magnitude of the different mechanisms of salt intrusion in the studied estuaries. Finally, we developed an equation, which computes the relative importance of the different components of the salt transport using only primary quantities, i.e. information about the geometry, discharge and tidal water variation at the mouth.
This equation reproduces the results from the model to a large extent, highlighting its potential applicability to e.g. assess the sensitivity of different types of salt transport in estuaries.

\section*{Open Research}

Software and data used in this study is made available online \cite{biem24_softw}.

%Open Research
% AGU requires an Availability Statement for the underlying data needed to understand, evaluate, and build upon the reported research at the time of peer review and publication.

%Additionally, authors should include an Availability Statement for the software that has a significant impact on the research. Details and templates are in the Availability Statement section of the Data & Software for Authors Guidance:
% https://www.agu.org/Publish-with-AGU/Publish/Author-Resources/Data-and-Software-for-Authors#availability

%For physical samples, use the IGSN persistent identifier, see the International Geo Sample Numbers section:
%https://www.agu.org/Publish-with-AGU/Publish/Author-Resources/Data-and-Software-for-Authors#IGSN
%%%%%%%%%%%%%%%%%%%%%%%%%%%%%%%%%%%%%%%%%%%%%%%

\acknowledgments
This work is part of the Perspectief Program SaltiSolutions, which is financed by NWO Domain Applied and Engineering Sciences (2022/TTW/01344701 P18-32 project5) in collaboration with private and public partners. We thank Elbert de Hon for his assistance creating the maps of the estuaries, Manuel D\'iez-Minguito for his explanation of the salinity behavior of the Guadalquivir Estuary, and Jiyong Lee for his contribution to our discussions.

%%%%%%%%%%%%%%%%%%%%%%%%%%%%%%%%%%%%%%%
\appendix

\section{Expression for the salt transport} \label{app:T}

The decompositions of the flow velocity, salinity and water level presented in Eq.~\ref{eq:dec} are inserted in the equation for salt transport Eq.~\ref{eq:Tdef}, which gives 
\begin{eqnarray} \label{eq:Tlong}
	T &=& (H + \eta_{st}) \bar u_{st} \bar s_{st} + (H + \eta_{st}) \left(\bar u_{ti} \bar s_{ti}\right)_{st} + (H + \eta_{st}) \overline{(u_{st}^\prime s_{st}^\prime)} + (H + \eta_{st}) \overline{(u_{ti}^\prime s_{ti}^\prime)}_{st} \nonumber \\
	&+& \bar u_{st} (\eta_{ti} \bar s_{ti})_{st} + \bar s_{st} (\eta_{ti} \bar u_{ti})_{st} + (\eta_{ti} \bar u_{ti} \bar{s}_{ti})_{st} 
	+\overline{u'_{st} (\eta_{ti} s'_{ti})}_{st} + \overline{s'_{st} (\eta_{ti} u'_{ti})}_{st} + (\eta_{ti} \overline{u'_{ti} s'_{ti}})_{st}  \nonumber \\
	&-& K_h (H+\eta_{st})\frac{\partial \bar s_{st}}{\partial x} - K_h (\eta_{ti} \frac{\partial \bar s_{ti}}{\partial x} )_{st} .
\end{eqnarray}
The term $\bar s_{st} (\eta_{ti} \bar u_{ti})_{st} $ is neglected in the following, as $(\eta_{ti} \bar u_{ti})_{st}$, the Stokes transport of water, is compensated by an equally large Stokes return flow, which exactly compensates for the salt transport by this term. 
To simplify Eq.~\ref{eq:Tlong}, we assume variations in water level are small with respect to the mean water depth, the variations in salinity in the tidal cycle are small with respect to the subtidal salinity, and the tidal flow is strong with respect to the subtidal flow. This is expressed mathematically as 
\begin{eqnarray} \label{eq:appro}
	&&[\eta_{ti} ]= [\varepsilon _1H], \quad  	
	[\eta_{st} ]= [\varepsilon_2 H], \quad  
	[\bar s_{ti}] = [\varepsilon_1 \bar s_{st}], \nonumber \\
	&&[s'_{ti}] = [\varepsilon_1 s'_{st}], \quad 
[\bar u_{st}] = [\varepsilon_3 \bar u_{ti}], \quad 	
	[u'_{st}] = [\varepsilon_3 u'_{ti}], %\quad 
	%\text{with} \quad  \varepsilon << 1,
\end{eqnarray}
in which $[\cdot]$ indicates the order of magnitude of a quantity, and $\varepsilon_1$, $\varepsilon_2$, and $\varepsilon_3$ are small parameters (much smaller than one). Inserting this  in Eq.~\ref{eq:Tlong} and retaining only the leading order terms yields Eq.~\ref{eq:Tdec}, and one extra term $\overline{s'_{st} (\eta_{ti} u'_{ti})}_{st}$. From the scaling analysis it is not clear that the latter is small, but a posteriori calculations from the model results indicated that this term has a maximum relative contribution of 7\%, which justifies neglecting it.

\section{Derivation of the flow velocity components} \label{app:vel}

The subtidal hydrodynamics are similar to what is used in \citeA{MC04} and the tidal hydrodynamics are described by \citeA{ianniello1979}, which is also the formulation used by \citeA{wei2016}. Differences concern the fact that we deal with connected channel segments, and the use of a different bottom boundary condition. 

For the subtidal hydrodynamics, we solve Eq.~\ref{eq:hyst} in the main text. Note that, to derive this equation (and also Eq.~\ref{eq:hyti}), the approximations presented in Eq.~\ref{eq:appro} are employed. 
The conditions at the horizontal boundaries read 
\begin{subequations} \label{eq:hbc}
	\begin{eqnarray}
		b \int_{-H}^{0} u_{st} dz &=& Q \quad \text{(river boundary)} , \\
		 \int_{-H_j}^{0} u_{j,st} dz &=&  \int_{-H_{j+1}}^{0} u_{j+1,st} dz \quad \text{(segment boundary)} , 
	\end{eqnarray}
\end{subequations}
where $j$ indicates the index of the segment. Note that, here and in the following, the vertical integrals are evaluated at $z=0$, the mean water level, and not at $z=\eta$, the true water level, because of approximations presented in Eq.~\ref{eq:appro}. The condition at the segment boundaries implies continuity of water transport. The vertical boundary conditions are
\begin{subequations}
	\begin{eqnarray}
		A_{v,st} \frac{\partial u_{st}}{\partial z}&=& 0 \quad \text{and} \quad w_{st} = 0 \quad \text{(water surface $z=0$)}, \\
		A_{v,st} \frac{\partial u_{st}}{\partial z}&=& S_{f,st} u_{st} \quad \text{and} \quad w_{st} = 0 \quad \text{(bottom $z=-H$)}.
	\end{eqnarray}
\end{subequations}
Their physical interpretation is that we set free-slip at the surface and partial slip at the bottom (see Eq.~\ref{eq:slip}) for horizontal velocity, and kinematic conditions for vertical velocity. 
The solutions follow from straightforward integration over $z$ and read
\begin{subequations} \label{eq:vel}
	\begin{eqnarray}
		\frac{\partial \eta_{st}}{\partial x} &=& -\frac{9}{20} \beta H \frac{\partial \bar{s}_{st}}{\partial x} - \frac{6}{5} \frac{A_{v,st} Q}{g b H^3}, \label{eq:eta}\\
		\bar u_{st} &=& \frac{Q}{b H},  \\		
		u'_{st} &=& \frac{Q}{b H} P_1(\tilde z) +\frac{g \beta H^3}{48 A_{v,st}}  \frac{\partial \bar{s}_{st}}{\partial x} P_2(\tilde z) , \label{eq:uu} \\
		w_{st} &=& - \frac{g \beta H^4}{48 A_{v,st}} \Big(\frac{\partial^2 \bar{s}_{st}}{\partial x^2} + L_b^{-1} \frac{\partial \bar{s}_{st}}{\partial x}\Big) P_w(\tilde{z}). 
	\end{eqnarray}
\end{subequations}
Here, $\tilde z = \frac{z}{H}$ and $P_1(\tilde z)$, $P_2(\tilde z)$ and $P_1(\tilde z)$ are polynomials, that describe the vertical structure of the currents and read
\begin{subequations} \label{eq:poly}
	\begin{eqnarray}
	P_1(\tilde z) &=& (\frac{1}{5} - \frac{3}{5} \frac{z^2}{H^2}) , \\
	P_2(\tilde z) &=&\Big(\frac{8}{5} - \frac{54}{5} \frac{z^2}{H^2} - 8 \frac{z^3}{H^3}\Big) ,\\
	P_w(\tilde z) &=& \Big(  2 \frac{z^4}{H^4} + \frac{18}{5} \frac{z^3}{H^3} - \frac{8}{5} \frac{z}{H}\Big).
	\end{eqnarray}
\end{subequations}
The difference with \citeA{MC04} concerns the use of partial slip condition Eq.~\ref{eq:slip} instead of a no-slip bottom boundary condition. 

%%%%%%%%%%%%%%%%%%%%

The equations of tidal motion are given by Eq.~\ref{eq:hyti}. Horizontal boundary conditions read 
\begin{subequations} \label{eq:hbct}
	\begin{eqnarray}
		b \int_{-H}^{0} u_{ti} dz &=& 0 \quad \text{(river boundary)} , \\
		\int_{-H_j}^{0} u_{j,ti} dz &=&  \int_{-H_{j+1}}^{0} u_{j+1,ti} dz \quad \text{(segment boundary)} ,\\
		 \eta_{ti,j} &=&  \eta_{ti,j+1} \quad \text{(segment boundary)} , \\
		 \eta_{ti} &=& A \cos(\omega t) \quad \text{(sea boundary)}, \label{eq:bc4}
	\end{eqnarray}
\end{subequations}
in which $A$ is the amplitude of the tidal water level at the estuary mouth, and $\omega$ is the angular frequency of the tidal constituent that is considered. Due to the inclusion of a segment which represents the adjacent sea, the boundary of the computational domain is not the sea boundary of the estuary. To impose condition \ref{eq:bc4}, the condition at the boundary of the computational domain is chosen in such a way that condition \ref{eq:bc4} is satisfied. Vertical boundary conditions are 
\begin{subequations}
	\begin{eqnarray}
		A_{v,ti} \frac{\partial u_{ti}}{\partial z}&=& 0 \quad \text{and} \quad w_{ti} = \frac{\partial \eta_{ti}}{\partial t} \quad \text{(water surface $z=0$)}, \\
		A_{v,ti} \frac{\partial u_{ti}}{\partial z}&=& S_{f,ti} u_{ti} \quad \text{and} \quad w_{ti} = 0 \quad \text{(bottom $z=-H$)}.
	\end{eqnarray}
\end{subequations}
Analytical solutions for this set of equations exists, and read for each segment
\begin{subequations} \label{eq:vels}
	\begin{eqnarray}
		\eta_{ti}, u_{ti},w_{ti} &=& \Re \{(\hat \eta_{ti}, \hat u_{ti}, \hat w_{ti})\exp(-i \omega t) \} , \label{eq:imag}\\
		\hat \eta_{ti} &=& \exp \left(-\frac{(x+L)}{2 L_b} \right) \left(C_1 \exp(k(x+L)) + C_2 \exp(-k(x+L)) \right) , \\
		\hat{ \bar{u}}_{ti} &=& \frac{g}{i \omega} \frac{\partial \hat \eta_{ti}}{\partial x} \left(\frac{B}{\delta_A} \sinh(\delta_A) -1 \right) ,\\	
		\hat u'_{ti} &=& \frac{g}{i \omega} \frac{\partial \hat \eta_{ti}}{\partial x} \left(B \cosh(\delta_A \frac{z}{H}) - \frac{B}{\delta_A} \sinh(\delta_A) \right),\\
		\hat w_{ti} &=& i\omega \hat \eta_{ti} - \frac{g}{i \omega} \left(\frac{d^2 \hat \eta_{ti}}{d x^2} + \frac{1}{L_b} \frac{d \hat  \eta_{ti}}{d x}\right) \left(\frac{B H}{\delta_A} \sinh(\delta_A \frac{z}{H}) - z\right) , 
	\end{eqnarray}
\end{subequations}
where $\Re$ denotes the real part of a complex variable and $C_1$ and $C_2$ are determined by the boundary conditions. The other parameters are defined as 
\begin{subequations} 
	\begin{eqnarray}
		&&\delta_A = \frac{(1+i)H}{\sqrt{\frac{2 A_{v,ti} }{\omega}}}, \\
		&&B = \Big( \cosh(\delta_A) + \frac{\delta_A}{2}  \sinh(\delta_A) \Big) ^{-1} , \\
		&&k = \sqrt{\frac{1}{(2 L_b)^2} + \frac{\omega^2}{g H} \left(\frac{B}{\delta_A \sinh(\delta_A)}-1 \right)^{-1}}.
	\end{eqnarray}
\end{subequations}
The physical interpretation of these parameters is that $|\delta_A|$ is the reciprocal Stokes number, where the latter measures the ratio of the thickness of the frictional boundary and water depth, and $k$ is the complex wavenumber. 

%%%%%%%%%%%%%%%%%%%%%%%%%%%%%%%%%%%%%%%%%%%%%%%%%%%%%%%

\section{Derivation of the expressions for salinity } \label{app:sal}

The starting point of the equations for salinity is the width-averaged advection-diffusion equation for salinity Eq.~\ref{eq:sb}.
In the following, we will decompose this equation, such that we get equations for the different components of salinity in Eq.~\ref{eq:dec}. The equation for tidal salinity is simplified, such that a semi-analytical solution of the salinity equations is possible. 
 To find an equation for the evolution of subtidal salinity, we substitute $\Phi = \Phi_{st} + \Phi_{ti}$, with $\Phi = (u,w,s,K_h,K_v)$, and subsequently perform an average over the tidal cycle. 
This gives 
\begin{equation} \label{eq:sst}
	\frac{\partial s_{st}}{\partial t} + u_{st} \frac{\partial s_{st}}{\partial x} + (u_{ti} \frac{\partial s_{ti}}{\partial x})_{st} + w_{st} \frac{\partial s_{st}}{\partial z} + (w_{ti} \frac{\partial s_{ti}}{\partial z})_{st} = \frac{1}{b} \frac{\partial}{\partial x} \Big(b K_{h,st} \frac{\partial s_{st}}{\partial x} \Big) + \frac{\partial}{\partial z} (K_{v,st}\frac{\partial s_{st}}{\partial z}).
\end{equation}
To proceed, we split salinity and horizontal velocity in a depth-averaged and a depth-perturbed part, indicated with a bar and a prime, respectively, and take the average over the depth afterwards. This yields for the depth-averaged subtidal salinity
\begin{equation} \label{eq:sstb}
	\frac{\partial \bar s_{st}}{\partial t} 
	+ \bar u_{st} \frac{\partial \bar s_{st}}{\partial x} 
	+ \overline{u'_{st} \frac{\partial s'_{st}}{\partial x}} 
%	+ \overline{w_{st} \frac{\partial s'_{st}}{\partial z}}
	+ (\bar u_{ti} \frac{\partial \bar s_{ti}}{\partial x})_{st} 
	+ (\overline{u'_{ti} \frac{\partial s'_{ti}}{\partial x}} )_{st} 
%	+ (\overline{w_{ti} \frac{\partial s'_{ti}}{\partial z}})_{st}
	= \frac{1}{b} \frac{\partial}{\partial x} (b K_{h,st} \frac{\partial \bar s_{st}}{\partial x})  , 
\end{equation}
where the no-flux boundary conditions at the bottom and surface and approximation Eq.~\ref{eq:appro} were used. 
We construct the depth-perturbed equation by subtracting the two previous equations from each other and get
\begin{multline} \label{eq:sstp}
	\frac{\partial s'_{st}}{\partial t}  
	+\bar u_{st} \frac{\partial s'_{st}}{\partial x}   
	+u'_{st} \frac{\partial \bar s_{st}}{\partial x}  
	+u'_{st} \frac{\partial s'_{st}}{\partial x}  
	-\overline{u'_{st} \frac{\partial s'_{st}}{\partial x}}  
+	w_{st} \frac{\partial s'_{st}}{\partial z}  
% -   \overline{w_{st} \frac{\partial s'_{st}}{\partial z}}
+	(\bar u_{ti} \frac{\partial s'_{ti}}{\partial x})_{st}  
+	(u'_{ti} \frac{\partial \bar s_{ti}}{\partial x})_{st}  \\ 
+	(u'_{ti} \frac{\partial s'_{ti}}{\partial x})_{st }  
-	(\overline{u'_{ti} \frac{\partial s'_{ti}}{\partial x}})_{st}  
+	(w_{ti} \frac{\partial s'_{ti}}{\partial z})_{st} 
%-	(\overline{w_{ti} \frac{\partial s'_{ti}}{\partial z}})_{st} 
=	\frac{1}{b} \frac{\partial}{\partial x} (b K_{h,st} \frac{\partial s'_{st}}{\partial x})  
+	\frac{\partial }{\partial z} (K_{v,st} \frac{\partial s'_{st}}{\partial z}) .
\end{multline}
These expressions allow us to calculate the evolution of the two components of $s_{st}$, which still depend on $s_{ti}$. To find an expression for the latter, Eq.~\ref{eq:sst} is subtracted from Eq. \ref{eq:sb}. This yields
\begin{multline}
	\frac{\partial  s_{ti}}{\partial  t} + 
	u_{ti} \frac{\partial  s_{st}}{\partial  x} + 
	u_{st} \frac{\partial  s_{ti}}{\partial  x}  + 
	u_{ti} \frac{\partial  s_{ti}}{\partial  x} - 
	(u_{ti} \frac{\partial  s_{ti}}{\partial  x})_{st}  + 
	w_{ti} \frac{\partial  s_{st}}{\partial  z}  + 
	w_{st} \frac{\partial  s_{ti}}{\partial  z}   + 
	w_{ti} \frac{\partial  s_{ti}}{\partial  z} -
	(w_{ti} \frac{\partial  s_{ti}}{\partial  z})_{st}  \\ = 	
	\frac{1}{b }\frac{\partial }{\partial  x} (b K_h \frac{\partial  s_{ti}}{\partial  x})  + 	
	\frac{\partial }{\partial  z} (K_v \frac{\partial  s_{ti}}{\partial  z}) .
\end{multline}
Since the solutions for tidal velocities are written as harmonic series, it is natural to decompose the tidal salinity also in harmonic components, that is writing $s_{ti}$ as \\ \noindent \mbox{$s_{ti} = \sum_{p=1}^{M}\Re \left[\hat s_{ti,p} \exp(-i p \omega t) +c.c. \right]$}, and neglect the contributions other than those of the dominant tidal constituent, so $M=1$. 
This gives, after redefining $\hat s_{ti,1} = \hat s_{ti}$,
\begin{align} 
	- i \omega \hat s_{ti}  + \hat u_{ti} \frac{\partial  s_{st}}{\partial  x} + u_{st} \frac{\partial \hat s_{ti}}{\partial  x} + \hat w_{ti} \frac{\partial  s_{st}}{\partial  z} +  w_{st} \frac{\partial \hat s_{ti}}{\partial  z} = 
	\frac{1}{b}	\frac{\partial }{\partial  x} (b K_{h,ti} \frac{\partial \hat s_{ti}}{\partial  x}) +  \frac{\partial }{\partial  z} (K_{v,ti} \frac{\partial \hat s_{ti}}{\partial  z}).
\end{align}
We look for a first-order solution to this equation. For this, we follow the scaling from \citeA{wei2016}, to obtain the first-order balanc, which reads
\begin{align} \label{eq:sti}
	- i \omega \hat s_{ti}  + \hat u_{ti} \frac{\partial  \bar s_{st}}{\partial  x}  + \hat w_{ti} \frac{\partial  s'_{st}}{\partial  z}  =  \frac{\partial }{\partial  z} (K_{v,ti} \frac{\partial \hat s_{ti}}{\partial  z}),
\end{align}
where we have approximated $ \frac{\partial  s_{st}}{\partial  x} \approx \frac{\partial  \bar s_{st}}{\partial  x} $. Model simulations which do not make this approximation were also performed, but the difference in the salt transports was smaller than 5\%. 
The separation in depth-averaged and depth-perturbed tidal salinity is performed on the output of the model, to keep the solution procedure as simple as possible. 

\section{Solutions for salinity using the Pritchard balance  } \label{app:pri}

Here we present the analytical solutions for salinity under the \citeA{pritchard1954} approximation. Under this approximation, Eq. \ref{eq:sstp} reduces to 
\begin{equation}
u'_{st} \frac{\partial \bar s_{st}}{\partial x}  
= \frac{\partial }{\partial z} (K_{v,st} \frac{\partial s'_{st}}{\partial z}) .
\end{equation}
This equation has an analytical solution, which is derived in \citeA{hansen1966} and reads
\begin{eqnarray} \label{eq:sstP54}
&&s'_{st} = \frac{H^2}{K_{v,st}} \frac{\partial \bar s_{st}}{\partial x} \left( \bar u P_3(z) + \alpha \frac{\partial \bar s_{st}}{\partial x } P_4(z) \right) , \quad \text{with} \label{eq:ssta} \\
&&P_3(\tilde z) = -\frac{7}{300} + \frac{1}{10} \tilde z^2 - \frac{1}{20} \tilde z^4 , 
\quad P_4(\tilde z) =  -\frac{23}{150} + \frac{4}{5} \tilde z^2 - \frac{9}{10} \tilde z^4 - \frac{2}{5} \tilde z^5. \nonumber 
\end{eqnarray}
For the tidal salinity, we solve Eq. \ref{eq:sti}, using the above expression for $s'_{st}$:
\begin{align} \label{eq:sti9}
	i \omega \hat s_{ti}  +  \frac{\partial }{\partial  z} (K_{v,ti} \frac{\partial \hat s_{ti}}{\partial  z}) = -\hat u_{ti} \frac{\partial  \bar s_{st}}{\partial  x}  - \hat w_{ti} \frac{H^2}{K_{v,st}} \frac{\partial \bar s_{st}}{\partial x} \left( \bar u \frac{\partial P_3(\tilde z)}{\partial z} + \alpha \frac{\partial \bar s_{st}}{\partial x } \frac{\partial P_4(\tilde z)}{\partial z} \right) , 
\end{align}
The right-hand side of Eq.~\ref{eq:sti9} consists of different terms, which scale linearly with either $\frac{\partial \bar s_{st}}{\partial x}$ or $\left(\frac{\partial \bar s_{st}}{\partial x}\right)^2$. Since the equation is linear, solutions for $s_{ti}$ also scale linearly with these quantities. Therefore, the solution can be written as 
\begin{equation} \label{eq:stia}
	s_{ti} = c_Q \left( \frac{\partial \bar s_{st}}{\partial x}\right)  + c_E \left(\frac{\partial \bar s_{st}}{\partial x}\right)^2. 
\end{equation}
The expression for $s_{ti}$ is very lengthy and interested readers can get the expression on request from the authors.
For $\bar s_{st}$, no analytical solution is available and Eq. \ref{eq:sstb} is solved numerically.  

\section{Derivation of an approximate expression for $T_{\overline{ti}}$} \label{app:Tti}

To find an approximate expression for $T_{\overline{ti}}$, first Eq.~\ref{eq:bsti} is solved for the amplitude of the tidally varying salinity $\hat{ \bar {s}}_{ti}$ (see also \ref{app:sal}), which gives
\begin{equation}
	\hat {\bar {s}}_{ti} = - \frac{i}{\omega} \left(\hat {\bar {u}}_{ti}\frac{\partial \bar s_{st}}{\partial x} + \overline{ \hat w_{ti} \frac{\partial s'_{st}}{\partial z}}\right). 
\end{equation}
Using this, we write $T_{\overline{ti}}$ as 
\begin{equation} \label{eq:tti2}
T_{\overline{ti}} = \left(\bar u_{ti} \bar s_{ti}\right)_{st} = \frac{b H}{2 \omega} |\hat {\bar {u}}_{ti}| |\overline{ \hat w_{ti} \frac{\partial s'_{st}}{\partial z}}| \ \sin \left(\arg(\overline{ \hat w_{ti} \frac{\partial s'_{st}}{\partial z}}) - \arg(\hat {\bar {u}}_{ti}) \right). 
\end{equation}
To proceed, the vertical velocity is approximated as
\begin{equation}
	\hat w_{ti} = - \int_{-H}^z \frac{1}{b} \frac{\partial}{\partial x} \left(b \hat u_{ti} \right) dz' \approx - \int_{-H}^z \frac{1}{b} \frac{\partial}{\partial x} \left(b \hat {\bar{u}}_{ti} \right) dz' = - \frac{z+H}{H} \frac{\partial \hat \eta_{ti}}{\partial t}, 
\end{equation}
where we have used the continuity equation Eq.~\ref{eq:tc} for the first equality and the depth-averaged version of this equation for the last equality. The approximation of $w_{ti}$ is justified a posteriori by confirming that the difference between $T_{\overline{ti}}$ from Eq.~\ref{eq:tti} and from the full model differs by less than 20\% in the regions of interest.
With this, we find 
\begin{equation}
\overline{ \hat w_{ti} \frac{\partial s'_{st}}{\partial z}} \approx - i \frac{\omega}{H} \hat \eta_{ti} s'_{st} (z=0)
\end{equation} 
When substituting this expression in Eq.~\ref{eq:tti2}, and using Eq.~\ref{eq:sstP54} for $s'_{st} (z=0)$, we find Eq.~\ref{eq:tti}.

\section{Expressions of the length scales} \label{app:ls}

The expressions for the length scales in Eq.~\ref{eq:L} are given by 
\begin{subequations}  \label{eq:Ls}
	\begin{eqnarray} 
		L_D &=& \frac{K_{h,st}}{\bar u_{st} },\\
		L_{E1} &=& -\frac{H^2}{K_{v,st}} \bar u_{st} \overline{P_1(z) P_3(z)},\\
		L_{E2} &=& \Big(-\frac{g \beta s_{oc} H^5}{48 A_{v,st} K_{v,st}} (\overline{P_2(z) P_3(z)} + \overline{P_1(z) P_4(z)}) \Big)^\frac{1}{2},\\
		L_{E3} &=& \Big(- \frac{g^2 \beta^2 s_{oc}^2 H^8}{48^2 \bar u_{st} A_{v,st}^2 K_{v,st}} \overline{P_2(z) P_4(z)} \Big)^\frac{1}{3},\\
		L_{T1} &=& \frac{(\bar c_{Q} \bar u_{ti})_{st}}{\bar u_{st}},\\
		L_{T2} &=& \Big(\frac{s_{oc}(\bar c_{E} \bar u_{ti})_{st}}{\bar u_{st}}\Big)^\frac{1}{2} , \\
		L_{T3} &=& \frac{(\overline{c'_{Q} u'_{ti}})_{st}}{\bar u_{st}},\\
		L_{T4} &=& \Big(\frac{s_{oc}(\overline{c'_{E} u'_{ti}})_{st}}{\bar u_{st}} \Big)^\frac{1}{2} .
	\end{eqnarray}
\end{subequations} 
Here, $c_Q$ and $c_E$ are constants occurring in the expression for tidally varying salinity, that are defined in Eq.~\ref{eq:stia}.
%

%%%%%%%%%%%%%%%%%%%%%%%%%%%%%%%%%%%%%%%%%%%%%%%%%%%%%

%%% ------------------------------------------------------------------------ %%
%% References and Citations

%%%%%%%%%%%%%%%%%%%%%%%%%%%%%%%%%%%%%%%%%%%%%%%
%
% \bibliography{<name of your .bib file>} don't specify the file extension
%
% don't specify bibliographystyle

% In the References section, cite the data/software described in the Availability Statement (this includes primary and processed data used for your research). For details on data/software citation as well as examples, see the Data & Software Citation section of the Data & Software for Authors guidance
% https://www.agu.org/Publish-with-AGU/Publish/Author-Resources/Data-and-Software-for-Authors#citation

%%%%%%%%%%%%%%%%%%%%%%%%%%%%%%%%%%%%%%%%%%%%%%%

%% REFERENCES
%\bibliographystyle{copernicus}

\bibliography{literature}

%Reference citation instructions and examples:
%
% Please use ONLY \cite and \citeA for reference citations.
% \cite for parenthetical references
% ...as shown in recent studies (Simpson et al., 2019)
% \citeA for in-text citations
% ...Simpson et al. (2019) have shown...
%
%
%...as shown by \citeA{jskilby}.
%...as shown by \citeA{lewin76}, \citeA{carson86}, \citeA{bartoldy02}, and \citeA{rinaldi03}.
%...has been shown \cite{jskilbye}.
%...has been shown \cite{lewin76,carson86,bartoldy02,rinaldi03}.
%... \cite <i.e.>[]{lewin76,carson86,bartoldy02,rinaldi03}.
%...has been shown by \cite <e.g.,>[and others]{lewin76}.
%
% apacite uses < > for prenotes and [ ] for postnotes
% DO NOT use other cite commands (e.g., \citeA, \citep, \citeyear, \citealp, etc.).
% \nocite is okay to use to add references from your Supporting Information
%

\end{document}